\documentclass[letter,fleqn,twoside]{article}
\pdfoutput=1

\usepackage{amsmath,amssymb,amscd,graphicx,epsfig}
 \usepackage{espcrc2}
\readRCS
$Id: espcrc2.tex,v 1.2 2004/02/24 11:22:11 spepping Exp $
\usepackage[figuresright]{rotating}

\newcommand{\be}{\begin{equation}}
\newcommand{\ee}{\end{equation}}
\newcommand{\ba}{\begin{eqnarray}
\addtolength{\abovedisplayskip}{1.2mm}
\addtolength{\belowdisplayskip}{1.2mm}}
\newcommand{\ea}{\end{eqnarray}}
\newcommand{\bea}{\begin{eqnarray*}}
\newcommand{\eea}{\end{eqnarray*}}
\newcommand{\barr}{\begin{array}}
\newcommand{\earr}{\end{array}}

\newcommand{\FF}{{\mathsf F}}
\newcommand{\sss}{{\! s}}
\newcommand{\aaa}{{\mbox{\scriptsize \sc a}}}
\newcommand{\bbb}{{\mbox{\scriptsize \sc b}}}
\newcommand{\N}{{\mathcal{N}}}

\newcommand{\OO}{{\mathcal{O}}}

\newcommand{\Z}{{\mathbb{Z}}}

\newcommand{\C}{{\mathbb{C}}}
\newcommand{\PP}{{\mathbb{P}}}

\newcommand{\lb}{\label}

\newcommand{\ra}{\rightarrow}
\newcommand{\lra}{\longrightarrow}

\newcommand{\wt}{\widetilde}
\newcommand{\td}{\tilde}

\newcommand{\al}{\alpha}

\newcommand{\p}{\partial}
\newcommand{\dl}{\delta}

\newcommand{\ld}{\lambda}

\newcommand{\vp}{\varphi}

\newcommand{\om}{\omega}
\newcommand{\Om}{\Omega}

\newcommand{\ccap}{\cdot}

\newcommand{\Tr}{{\rm Tr}}
\newcommand{\ch}{{\rm ch}}
\newcommand{\rk}{{\rm rk}}
\newcommand{\vol}{{\rm vol}}
\newcommand{\Ext}{{\rm Ext}}
\newcommand{\Hom}{{\rm Hom}}

\title{D-branes at Singularities and String Phenomenology}

\author{Dmitry Malyshev, Herman Verlinde\address{
        Physics department, Princeton University, Princeton, NJ 08544}%
        \thanks{This work was supported by the National Science Foundation under grants
                PHY-0243680 and DMS-0606578, and
                by the Russian Foundation of Basic Research, grant RFBR 06-02-17383 (D.M.).}}

\runtitle{D-branes at Singularities%, Compactification, and Hypercharge
}
\runauthor{H. Verlinde}

\begin{document}

\begin{abstract}
In these notes we give an introduction to some of
the concepts involved in constructing SM-like gauge theories in
systems of branes at singularities of CY manifolds.
These notes are an expanded version of lectures given by Herman
Verlinde at the Cargese 2006 Summer School.

\end{abstract}

\newcommand{\ccc}{{\mbox{\large $c$}}}

\maketitle

\tableofcontents

\section{Introduction}

Over the past 10 years, fueled by the deepened understanding of duality and D-brane physics,
open string theory has evolved into an increasingly successful tool for building 4-d
supersymmetric field theories. In particular, it is now understood that by taking a judicious low energy
limit of the world-volume theory on a stack of D3-branes, one recovers a purely 3+1-dimensional
gauge theory, decoupled from gravity and all extra dimensional dynamics.
In this decoupling limit, the closed string background freezes into a set of non-dynamical, tunable
gauge invariant couplings. By placing one or more D-branes near
various types of geometric singularities, realizations of large classes of gauge theories have been
uncovered, and a detailed dictionary between geometric and gauge theory data is emerging.
Open string theory has become a preferred duality frame for representing weakly
coupled 3+1-d Quantum Field Theories in string theory.

A central characteristic of interacting Quantum Field Theories is that couplings and masses are not
constants but non-trivial functions of the energy scale, set by the relevant dynamical process, such as a
collision of two particles. %Thus the coupling depends on the dynamics of the fields.
This running behavior of the couplings has been beautifully explained by the Wilson renormalization group. In recent years, the renormalization group has been viewed from a fundamentally
new perspective via the embedding of QFT within string theory via D-branes.
String theory has no coupling constants. Instead all couplings are fields that
attain certain expectation values, set by solving their equation of motion.
In the holographic interpretation, the RG scale is one of the extra dimensions.
In this way the dependence of couplings on the energy scale
becomes a more intuitive dependence of a field on a spatial coordinate.

The relation between the RG scale and a coordinate in the
internal space lies at the heart of one of the most beautiful
advances in string theory, the AdS/CFT correspondence
\cite{Maldacena:1997} \cite{Gubser:1998} \cite{Witten:1998qj}
\cite{Morrison:1998cs}. The gauge invariant
couplings of a gauge theory, that has
been obtained as the low
energy limit of open string theory on a stack of D-branes,
are set by the closed string background in
which the D-branes are immersed. This background geometry is not fixed, but
dynamically influenced by presence of the D-branes. The backreaction
creates a warped local neighborhood, in which the distance from the branes
naturally becomes identified with an energy scale. This interpretation is
known as the holographic renormalization group.
A particularly well-studied and non-trivial example of this correspondence is
the Klebanov-Strassler solution \cite{Klebanov:1999rd} \cite{Klebanov:2000hb}
where the RG running of gauge couplings is
mapped to the logarithmic dependence of the B-field that solves the
supergravity equations of motion in the extra dimensions.

The gauge theory-gravity correspondence has been explored in
many ways, and an elaborate dictionary relating quantities in
the two dual systems is being uncovered. In these lectures we
will attempt to give a pedagogical introduction to the construction
of gauge theories from D-branes in string theory, and the correspondence
between the two sets of data. We will focus on D-branes placed at singularities of
Calabi-Yau manifolds in type IIB string theory.
We will assume that the D-branes span 3+1-d Minkowski space.
The gauge and matter fields arise from open string
modes on the D-branes, whereas parameters such as
gauge couplings, Yukawa coupling,  etc, correspond to the geometric properties
of the internal space.

Our main motivation for the study of D-branes at Calabi-Yau singularities is that it provides an interesting alternative route towards string phenomenology. As the first step in this program, one
needs to establish a sufficiently
general  dictionary between gauge theory quantities and local and global properties of singular CY
threefolds. Assuming one can succeed with this first step, it then becomes valid
to try to find explicit realizations of
world-volume gauge theories on D-branes  at singularities, that reproduce the  Standard Model of
particle physics. Moreover,  since every decoupled theory, via its space of tunable couplings,
stretches out over a sizable open neighborhood within the space of 4-d gauge
theories, one can even aim to reproduce the Standard Model spectrum and couplings
within their phenomenological bounds.
Though clearly a non-trivial  challenge, this question is still  much less ambitious,
and thus easier to answer, than finding a fully realistic closed string background
via the conventional top-down approach.
%Nonetheless, it would be a useful first step.

%The closed strings in Minkowski space correspond to dynamical fields,
%e.g. to the graviton.
%If a coupling also corresponds to closed string mode, why don't
%we see its propagation in our world?
%The answer is that there are two types of modes:
%there are modes that propagate through the entire compact manifold
%and there are modes localized near the D-branes.
%The modes of the first kind decouple from the fields on the brane:
%they are non dynamical fields
%represented by the parameters in the gauge theory.
%The localized modes survive as dynamical fields in the gauge theory,
%these modes will be crucial in cancelation of the mixed anomalies
%through the Green-Schwarz mechanism
%\cite{Green:1984sg}.

\bigskip

The organization of the lectures is as follows.

In section 2 we review some properties of the background Calabi-Yau
geometry.  We will specialize our discussion to
Calabi-Yau singularities that take the form of a complex cone over a
complex two-dimensional base space.
At the end of Section 2 we give a short description of the geometry
of del Pezzo surfaces, which will form our canonical choice for the
base of the Calabi-Yau singularity.
In these lectures, we will often refer to the canonical class
of complex manifolds.
%will be important in finding the chiral
%fields in the gauge theory and in identifying the anomalous $U(1)$
%gauge groups.
As a geometrical intermezzo,
we calculate the canonical class in some simple examples.

In section 3 we consider the D-branes placed at
the tip of the cone. We introduce the notion of fractional branes
\cite{Douglas:1996sw} \cite{Diaconescu:1997br},
which can be visualized as well-chosen stable
bound states of D-branes
\cite{Douglas:1995bn}\cite{Polchinski:1996na}.
Their world-volume gauge theory
supports non-trivial magnetic flux. As a result, the D-brane
carries charges of lower dimensional D-branes.
%The supersymmetric the D-branes should satisfy
%the BPS condition.
We introduce the central charge as
an invariant characteristic of BPS D-branes.
% which can be used to express various couplings of the
%world volume gauge theory.
A collection of fractional D-branes preserves
supersymmetry provided all their central charges have the same
phase
\cite{Douglas:2000gi}.

In section 4 we describe the quiver gauge theories
for the D3-brane at the tip of the cone.
We present simple formulas that count the number of
massless open string states at the intersection between the fractional branes.
In the gauge theory, these correspond to the matter fields charged under the gauge
groups that act on the two ends of the open string. As examples, we consider
branes near orbifold singularities and on (toric and non-toric) del Pezzo singularities.

In section 5 we discuss the relation between the parameters in the
quiver gauge theory and the closed string modes. We outline the computation
of the superpotential, and of the spectrum of massless $U(1)$ vector bosons.

Finally, after collecting all the necessary technology, we outline our general bottom-up
approach to string phenomenology in section 6. We apply the acquired insights to a
concrete construction of an SM-like theory,  based on a single D3-brane near
a suitably chosen del Pezzo 8 singularity. We specify a simple topological condition
on the compact  embedding of the $dP_8$ singularity, such that only hypercharge survives
as the massless gauge symmetry.
We summarize some of  our conclusions in section 7.

\medskip

\newcommand{\sump}{\mbox{\large $\sum\limits_{\raisebox{.5mm}{\scriptsize $p$}}$}}

\section{Calabi-Yau~cones~and~del~Pezzo~surfaces}

In this section we review some properties of Calabi-Yau and del Pezzo
manifolds.
%The presentation is intended to be more intuitive than
%mathematically strict.
We assume some basic knowledge about complex manifolds,
line bundles and divisors.
The description of these topics can be found, for example, in
\cite{Griffiths}.

\subsection{Calabi-Yau cones}

Calabi-Yau manifolds are complex Ricci flat manifolds that provide a
good background for string compactifications.
The Ricci flatness is necessary for the absence of conformal
anomalies.  Calabi-Yau manifolds admit one covariantly constant spinor, and
hence preserve at least one
supersymmetry
\cite{GSW}.%
\footnote{
In the presence of fluxes or branes the Ricci flatness condition may
be modified, also one can break the supersymmetry completely.}

More strictly, a Calabi-Yau manifold $Y$ is
a compact Kahler manifold with a vanishing first Chern class,
$c_1(Y)=0$.
We will assume that the manifold has three complex dimensions.
A complex manifold is called Kahler if its Kahler form
\be
J=ig_{\mu\bar\nu}dx^\mu\wedge dx^{\bar\nu}
\ee
is closed \cite{Griffiths}.
One of the properties of an $n$-dimensional
Calabi-Yau manifold is that it has a
nowhere vanishing holomorphic $n$-form $\Om$.
The form $\Om$ has only holomorphic indices and depends on
$z_1,\ldots,z_n$ (not $\bar z_1,\ldots,\bar z_n$)
\be
\Om=\Om_{1\ldots n}dz_1\ldots dz_n\, ; \qquad
\bar \p \Om=0\, .
\ee
For a general complex manifold $M$, $\Om$ may have zeros and poles.
The corresponding divisor is called the canonical class and
is denoted by $K(M)$.
%The divisor of $\Om$, the collection of its zeros, is
%the canonical class $K(Y)$.
We will use the same notation $K(M)$ for the line bundle associated
to this divisor.%
\footnote{
In general, the category of line bundles is equivalent to the
category of divisors, i.e. for every divisor there is a
corresponding line bundle and vice versa, the sum of two divisors is
the tensor product of the line bundles \cite{Griffiths}.}
%This category of the line bundles for a manifold $M$ is called the
%Piccard group of $M$, ${\rm Pic}(M)$.
The form $\Om$ can be considered as a section of the line bundle
$K(M)$.

For a general compact manifold $M$ the class of the canonical bundle is minus
the first Chern class of the holomorphic tangent bundle
\be
K(M)=-c_1(TM).
\ee
The existence of non zero section of $K(M)$
is equivalent to the triviality of
the bundle, i.e. $K=-c_1=0$, which coincides with the Calabi-Yau
condition.

We will use the triviality of the canonical class for the definition of
non compact Calabi-Yau manifolds.
The motivation is that both the existence of the covariantly
constant spinor $\eta$ and the existence of the
Ricci flat metric are related to
the existence of everywhere non zero holomorphic $n$-form $\Om$
\cite{GSW}.
%\be
%\Om_{1\ldots n}=\eta^\dag \G_{1\ldots n}\eta\, .
%\ee
Note, that the local Calabi-Yau condition is less restrictive than the global
one.\footnote{E.g. the complex projective space $\PP^1$ is locally a complex
line $\C^1$, the latter is evidently a non compact Calabi-Yau while
the former is not a Calabi-Yau.}
%Thus the first step of our program is to start with a simpler, i.e. local
%Calabi-Yau condition and find the quiver gauge theory that
%resembles the Standard Model.
%The second step is to find an example of a compact Calabi-Yau
%that has a local region obtained in step one.

%The manifolds we will use for the construction of the Standard Model will be singular.
A singular manifold will be called a Calabi-Yau if it can be obtained
by a complex or a Kahler deformation from a smooth Calabi-Yau.
For example, the singularity of the conifold can be either deformed or
resolved and in both cases the resulting manifold is a smooth CY
\cite{Candelas:1989js}.

A rich class of Calabi-Yau singularities is provided by
complex cones over a base space $X$ of complex dimension two.
In order to get a CY cone we, first, take
a line bundle over the
base space $X$ such that the canonical class of the total space of the
bundle is trivial and, second, shrink the zero section of the bundle
to a point. The line bundle over $X$ is the normal bundle to $X$
inside $Y$. We denote it by $N_X$.
The line bundle is not arbitrary: it is completely specified
by the condition of vanishing canonical class.
From the adjunction formula it follows that
the divisor for the line bundle $N_X$ is equal to the canonical class of
the base space.
Indeed, the maximal holomorphic $n$-form on $Y$ restricted to $X$
can be decomposed in an $(n-1)$-form on $X$ and a
one-form "perpendicular" to $X$
\be
K(Y)|_X=K(X)\otimes N^*_X,
\ee
and since from the Calabi-Yau condition it follows that the restriction
of the canonical class  to the base is trivial $K(Y)|_X=1$, we have
%where by $N^*_S$ we denote the space of one forms on $S$,
%$N_S\otimes N_S^*=1$.
\be\lb{CYcond}
N_X=K(X).
\ee
In the following we will consider a particular class of Calabi-Yau singularities of this type,
for which the base $X$ is a del Pezzo surface.

To specify the geometry of the CY cone,
let $ds_X^2 = h_{a\bar{b}}dz^a dz^{\bar{b}}$ be a K\"ahler-Einstein metric over
the base $X$ with $R_{a\bar{b}} = 6 h_{a\bar{b}}$ and first Chern class
$\omega_{a\bar{b}} = 6 i R_{a\bar{b}}$.
Introduce the one-form $\eta = {1\over 3} d\psi + \sigma$ where $\sigma$ is defined
by $d\sigma = 2\omega$ and $0<\psi<2\pi$ is the angular coordinate for a circle bundle
over the base $X$. Then the Calabi-Yau metric can then be written as follows
\be
\label{one}
ds^2_{Y} = dr^2 + r^2\eta^2 + r^2 ds^2_X %h_{a\bar{b}} dz^a dz^{\bar{b}} .
\ee
For the non-compact cone, the $r$-coordinate has infinite range.
Alternatively, we can think of the cone as a localized region
within a compact CY manifold, with $r$ being the local radial coordinate distance. % from the singularity.
%We will consider both cases.

\newcommand{\qq}{{\mathsf q}}
\newcommand{\pp}{{\mathsf p}}
\newcommand{\rr}{{\bf  r}}

%${}$

\bigskip

 {\small
\addtolength{\baselineskip}{-.2mm}

\subsection{\small Intermezzo: the canonical class}

The canonical class is an important geometric characteristic that
will show up in many places through out these lectures.
%In particular, in the analysis of the anomalous $U(1)$ gauge groups.
Let us give some examples of how one can calculate it.
In the following $\PP^n$ will denote the complex $n$-dimensional projective
space.

\bigskip

\noindent
{\bf Example 1. Canonical class of $\PP^n$.}

The projective plane $\PP^1$ has two coordinate charts parameterized by $(1,z_1)$ and
 $(z_0,1)$.  The two charts are glued together via
$z_1=z_0^{-1}$.
Line bundles over $\PP^1$ are classified by their divisor.
A section of $\OO(n)$ bundle for $n>0$ may have $n$
zeros and no poles or $n+1$ zeros and a single pole etc.

Let $\ld_0=1$ be a section of $\OO(n)$ bundle in the chart
$(1,z_1)$, then in the chart $(z_0,1)$ the section is $\ld_1=z_0^n$.
The transition function is therefore
\be
\label{trans}
f_{0\ra 1}=\ld_1/\ld_0=(z_0/z_1)^n.
\ee
%the transformation of the generic section is
%\be\lb{trans}
%\ld_1=f_{0\ra 1}\cdot\ld_0=\frac{z_0^n}{z_1^n}\ld_0
%\ee
The sections of the canonical bundle on $\PP^1$ are holomorphic
one forms.
Consider the one-form $\Om\!=\!dz_1$ in the chart $z_0\neq 0$.
On the intersection $\Om\!=\!-z_0^{-2}dz_0$.
The one-form  $\Om$ thus has a double pole at $z_0=0$, and the
canonical class is therefore $K(\PP^1)={\cal O}(-2)$.

For $\PP^n$, any section of the canonical bundle is a
holomorphic $n$-form. Consider the chart $z_0\neq 0$ with coordinates
$(1,z_1,\ldots,z_n)$ and the form
\be
\Om=dz_1\ldots dz_n.
\ee
In the chart $z_1\neq 0$ introduce the coordinates
\be
(z_1,z_2,\ldots,z_n)
=(\frac{1}{y_1},\frac{y_2}{y_1},\ldots,\frac{y_n}{y_1})
\ee
then the holomorphic form reads
\be
\Om=-\frac{1}{y_1^{n+1}}dy_1\ldots dy_n.
\ee
Note that in homogeneous coordinates $y_1\!=\!z_0/z_1$.  Consequently,
the $(n+1)$-th order pole at $y_1=0$ corresponds to the pole at the
hyperplane $z_0=0$.
Thus the canonical bundle
\be
K(\PP^n)=-(n+1)H,
\ee
where $H$ is the hyperplane class of $\PP^n$.

We can reach the same conclusion by using the total Chern
class of $\PP^n$, given by
\be
c(\PP^n)=(1+H)^{n+1} %=1+(n+1)H+\ldots
\ee
with $H$ the hyperplane class.
The canonical class is therefore
\be
K(\PP^n)\!=\!-c_1\!=\!-(n+1)H,
\ee
as before. %The Chern class is a useful characteristic for compact manifolds.
%Though it is not as convenient in the non compact case
%where it's easier to use the canonical class.

\bigskip

\noindent
{\bf Example 2. Line bundle ${\cal O}(n)$ over $\PP^1$.}

Let $\lambda$ be a section of the ${\cal O}(n)$ line bundle
over $\PP^1$.
Denote by $\lambda_0$ and $\lambda_1$ the restrictions of $\ld$
to the two charts $(1,z_1)$ and $(z_0,1)$.
 Let us consider the section of the canonical bundle for the total
space of the line bundle in the chart $z_1=1$
\be
\Om=dz_0d\ld_1
\ee
In the intersection of the two charts, we have
\be
\Om=d(\frac{z_0}{z_1})d(\frac{z_0^n}{z_1^n}\ld_0)
\ee
so that in the chart $z_0=1$
\be\lb{canoo}
\Om=-\frac{1}{z_1^{n+2}}dz_1d\ld_0
\ee
For $n=-2$, $\Om$ has a non zero section,
consequently the total space of the $\OO(-2)$ line bundle has a
vanishing canonical class.%
%Hence the canonical class is trivial for the ${\cal O}(-2)$ bundle over $\PP^1$.
\footnote{
If we interpret the $\OO(-2)$ bundle as the normal bundle to $\PP^1$
inside the total space, then $N_{\PP^1}=\OO(-2)=K(\PP^1)$,
in accordance with the general formula (\ref{CYcond}).}

}

\medskip

{\small
\bigskip

\noindent
{\bf Example 3. $\C^2$ blown up at the origin.}

The blowup of $\C^2$ at the origin is
%a new manifold,denoted by $\wt{\C}^2$. It has a $\PP^1$
obtained by inserting a $\PP^1$
instead of the point at the origin of $\C^2$.
The blowup of $\C^2$ will be denoted by $\wt{\C}^2$.
The blown up $\PP^1$ is called the exceptional divisor and is
denoted by $E$.

Near the exceptional divisor,
the coordinates on $\wt{\C}^2$ can be written in the form
\be\lb{blowup}
z_0=\ld w_0, \;\;\;\;z_1=\ld w_1
\ee
where $(w_0,w_1)$ are homogeneous coordinates on $\PP^1$ and $\ld$
is the radial direction.
In order for this parametrization to be compatible with the
projective invariance of $\PP^1$, $\ld$ should be a section of
$\OO(-1)$ bundle over $\PP^1$ \cite{Griffiths}.

Before the blowup, $\C^2$ has the following
holomorphic two-form
\be\lb{cc1}
\Om=dz_0dz_1
\ee
After the blowup, this form is modified
in the vicinity of the exceptional divisor.
Using the coordinates (\ref{blowup}), we get in the chart $w_1=1$
\be\lb{cc2}
\Om=d (\ld_1 w_0) d (\ld_1 w_1)=\ld_1 dw_0 d\ld_1
\ee
The form $\Om$ has a zero at $\ld=0$, i.e. at the
exceptional divisor $E$.
Consequently the canonical class of $\wt{\C}^2$ is
\be
K(\wt{\C}^2)=E.
\ee
We will use this result in the next subsection.
}
\addtolength{\baselineskip}{.2mm}
\bigskip

\subsection{Del Pezzo surfaces}

The mathematical definition of del Pezzo surface is rather
abstract: it is a two-dimensional complex manifold such
that its anticanonical bundle is ample.
We will simply use the fact that any dell Pezzo surface is either a
$\PP^1\times\PP^1$ or a $\PP^2$ blown up at $n$ points,
where $0\geq n\geq 8$.
%, i.e. its
%divisor has positive intersections with all effective
%divisors on the surface.\footnote{
%Recall that any effective divisor is a sum of analytic hypersurfaces
%with positive coefficients. A hypersurface is called analytic if on
%every coordinate patch it is defined by an equation.}
%In particular, the anticanonical divisor
%should have positive  self intersection. In practice, del Pezzo surfaces may be obtained
%by blowing up $n$ points on $\PP^1\times\PP^1$ or $\PP^2$.  We will
%mostly work with the latter.
The $\PP^2$ blown up at $n$ points will be called the $n$-th del Pezzo surface
and denoted by $dP_n$.

Locally,
%canonical divisor will have the same form as for the $\wt \C^2$, i.e.
every blown up point supports a $\PP^1$, called an exceptional divisor
$E_i$. % Another divisor is the hyperplane class $H$ on $\PP^2$.
%Every exceptional divisor $E_i$ is an independent cycle on $dP_n$.
Together with the hyperplane class $H$ of $\PP^2$ the exceptional
divisors form a basis for $H_2(dP_n,\Z)$.
Thus the dimension of $H_2(dP_n,\Z)$ is
\be
b_2(dP_n)=n+1.
\ee
Also there is a class of a point, i.e. the zero cycle, and the
class of the four-cycle: %, corresponding to $\C^2\subset \PP^2$
\be
b_0(dP_n)=b_4(dP_n)=1.
\ee
There are no other cycles thus the Euler characteristic of $dP_n$ is
\be
\chi(dP_n)=n+3
\ee
A del Pezzo surface is complex, so we can define complex cohomology classes $h^{p,q}$.
%From the Hodge star duality, $h^{p,q}=h^{2-p,2-q}$,
%it follows that the cohomologies of $X$ have the form
%\be
%\barr{ccccc}
%{}&{}&1&{}&{}\\{}&{0}&{}&{0}&{}\\
%{h^{2,0}}&{}&{h^{1,1}}&{}&{h^{0,2}}\\
%{}&{0}&{}&{0}&{}\\{}&{}&1&{}&{}\\\earr \ee
The corresponding dimensions are
\be
h^{(0,0)}=1,\;\;\;\;h^{(1,1)}=n+1,\;\;\;\;h^{(2,2)}=1.
\ee
The other cohomologies are trivial.

These divisors have the following intersections
\be
\barr{l}
H\cdot H=1\\[1mm]
H\cdot E_i=0\\[1mm]
E_i\cdot E_j=-\dl_{ij}
\earr
\ee
Note, that the exceptional divisors have negative self intersection,
$E\cdot E=-1$.
%In example 3 above we have shown that
The surface near the exceptional
$\PP^1$ has the form of $\OO_{\PP^1}(-1)$ bundle.
The self intersection of $E$ is defined as the intersection of $E$ with
a little perturbation of $E$ along the normal directions, which is equal
to the intersection of a section of $\OO_{\PP^1}(-1)$ with the zero
section.
The sign of the intersection depends on the relative orientation of
the two curves.
If a section has a zero, then the intersection is
$+1$, if it has a pole, the intersection is $-1$.

%Each exceptional divisor contributes to the canonical class.
In the examples
above we have shown that the canonical class of $\PP^2$ is $-3H$ and
every exceptional divisor contributes $E_i$, consequently
the canonical class of the $dP_n$ surface is
\be
K(dP_n)=-3H+E_1+\ldots+E_n.
\ee
%where $-3H$ comes from the canonical class of the $\PP^2$.
The self intersection of anti-canonical divisor is
\be
(-K)\cdot (-K) %=9 H\cdot H + E_1\cdot E_1+\ldots+E_n\cdot E_n
=9-n.
\ee
A necessary condition for a surface to be dell Pezzo is that its
anti-canonical divisor has positive self-intersection, i.e.
$n\leq 8$.

For $n\geq 3$
we will usually use the following basis of the two-cycles on $dP_n$
\be
\barr{lll}
K&=&-3H+\sum_{i=1}^n E_i\\[2mm]
\al_i&=&E_{i}-E_{i+1},\;\;\;\;\;\;\;\;\;
 i=1,\ldots,n-1\\[2mm]
\al_n&=&H-E_1-E_2-E_3
\earr
\ee
The intersection matrix in this new basis takes the form
\be
K \ccap \alpha_i = 0\\
\alpha_i \ccap \alpha_j = - A_{ij}
\ee
where $A_{ij}$ equals  minus the Cartan
matrix of the corresponding
Lie groups $E_3\equiv A_2\times A_1$,
$E_4\equiv A_4$, $E_5\equiv D_5$, and the exceptional groups $E_6$,
$E_7$ and $E_8$.

%The canonical class has self-intersection $9-n$.
%Note, that for $n=3,\ldots, 8$
%the intersection matrix for the cycles $\al_i$ on $dP_n$
%is minus the Cartan matrix for the Lie algebra $E_n$.

The complex structure of del Pezzo surfaces
depends on the position of the points of $\PP^2$
that we blow up, i.e. for $dP_n$ we have $2n$ complex parameters,
since every point has two complex coordinates.
Two surfaces are equivalent if they are related by an
automorphism of $\PP^2$, i.e. by a $PGl(3)$ transformation.
This group has
$3^2-1=8$ complex parameters, and so four generic points on
$\PP^2$ can be moved to given positions by $PGl(3)$.
The remaining coordinates of the blown up points parameterize the
non equivalent complex structures.
For $n>4$, the complex structure of $dP_n$ is parameterized by $2n-8$
parameters.

Some equations that define the embedding of the $dP_n$ surfaces in
the (products of) weighted projective
spaces can be found in \cite{Hubsch} \cite{Friedman:1997yq}.

\bigskip

{\small

\noindent
{\bf Example: del Pezzo 8 surface}

As an example, let us prove that an equation of degree 6 in the
weighted projective space $W\PP^3_{1123}$ is the $dP_8$ surface.
%More details on this type of calculations can be found in
%\cite{Hubsch}.
Recal that the weighted projective space $W\PP^3_{1123}$ consists of the points
$(u,v,x,y)\neq(0,0,0,0)$ up to identifications
$(u,v,x,y)\!\sim\! (\ld u,\ld v,\ld^2 x,\ld^3 y)$ with $\ld\neq 0$.
Let $X$ denote the surface defined by the homogeneous equation
of degree $6$ on $W\PP^3_{1123}$
\be\lb{dp8eq}
ay^2=bx^3+cv^6+du^6+\ldots
\ee
Let us verify that this surface has the same cohomology and space of deformations
as $dP_8$.

The total Chern class of the weighted projective space is
\bea
c(W\PP_{1123})=(1+H)^2(1+2H)(1+3H)
\eea
where $H$ is the hyperplane class in $W\PP_{1123}$.
The class, Poincare dual to the surface defined by an equation of
degree $6$ in $W\PP_{1123}$, is $6H$.
Consequently the total Chern class of
this surface is \cite{Hubsch}
\bea
c(X)=\frac{(1+H)^2(1+2H)(1+3H)}{1+6H}
\eea
Expanding the fraction and using the relation $H^3=0$ on $X$ we find
\bea
c(X)=1+H+11H^2
\eea
hence $c_1(X)=H$ and $c_2(X)=11 H^2$.
Since $6H$ is the Poincare dual two-form for $X\subset W\PP_{1123}$,
we can extend the integration over $X$ to the integration over
$W\PP_{1123}$. One obtains\footnote{
The factor $1/6$ comes from the fact that $\PP^3$ is a six-fold
cover of $W\PP_{1123}$, since the weighted projective space has a
$Z_2$ orbifold singularity near $(0,0,1,0)$:
$(u,v,1,y)\sim (-u,-v,1,-y)$;
and a
$Z_3$ orbifold singularity at $(0,0,0,1)$:
$(u,v,x,1)\sim (\om u,\om v,\om^2 x, 1)$ where $\om^3=1$.
}
\bea
\chi(X)&=&\int_X c_2(X)
=\int_{W\PP_{1123}} c_2(X)\wedge 6H\\
&=&\frac{1}{6}6\cdot 11=11
\eea
The second cohomology for $X$ is therefore
\bea
h^{1,1}=\chi-2=9,
\eea
in accord with the identification of the surface as $dP_8$.

%Let us now show that the equations of degree 6 in $W\PP_{1123}$ give
%all possible $dP_8$ surfaces.
As we have shown the number of parameters describing the $dP_8$
surfaces is $2n-8=8$.
The number of coefficient in equation
(\ref{dp8eq}) is $23$.
Let us show that $W\PP_{1123}$ has $15$ coordinate
reparameterizations.
The coordinate $y$ has degree 3, hence a generic change of
coordinates involves addition of a polynomial of degree 3 in $u$,$v$ and $x$
to $y$. %but we cannot add y to the other coordinates.
The transformations of $y$ together with its
rescaling give 7 parameters.
The coordinate $x$ has degree two, hence we can add a polynomial of
degree 2 in $u$ and $v$.
The transformations of $x$ give 4 parameters.
Also there are arbitrary $Gl(2)$ transformations of $u$ and $v$, which
give the last 4 reparameterizations.
Consequently, the
equations up to change of coordinates are described by $8$
parameters and any generic $dP_8$ surface can be described by some
equation.\footnote{
In this respect the del Pezzo surfaces are different from the K3
surface, because not all the K3 surfaces are algebraic
\cite{Aspinwall:1996mn}.}
%i.e. can be embedded by a system of equations in a projective space.
%The 8 parameters describing the $dP_8$ surfaces can be
%chosen to be the coordinates on the weighted projective space
%$W\PP^8_{122334456}$, see \cite{Friedman:1997yq}.

At special values of the complex structure moduli, the del Pezzo 8 surface may
form ADE type singularities, at which one or more of the %degree zero
2-cycles $\alpha_i$ degenerate. We will make use of this possibility later on.}

\smallskip

\section{D-branes on a Calabi-Yau singularity}

The del Pezzo surface $X$ forms a four-cycle within the
full three-manifold $Y$, and itself supports several non-trivial
two-cycles.
%\footnote{We will give a more detailed description of
%del Pezzo surfaces and their 2-cycles in section 3.}
Now, if we consider IIB string theory on a del Pezzo
singularity, we should expect to find a basis
of  D-branes that spans the complete
homology of $X$: the del Pezzo 4-cycle itself may be
wrapped by any number of D7-branes, any 2-cycle within $X$
may be wrapped by one or more D5 branes, and the point-like
D3-branes occupy the 0-cycle within $X$.
We will now summarize some of their properties.

 \medskip

\begin{figure}[hbtp]
\begin{center}
%\leavevmode\hbox{\epsfxsize=7.5cm \epsffile{East3.pdf}}\\[3mm]
\epsfig{figure=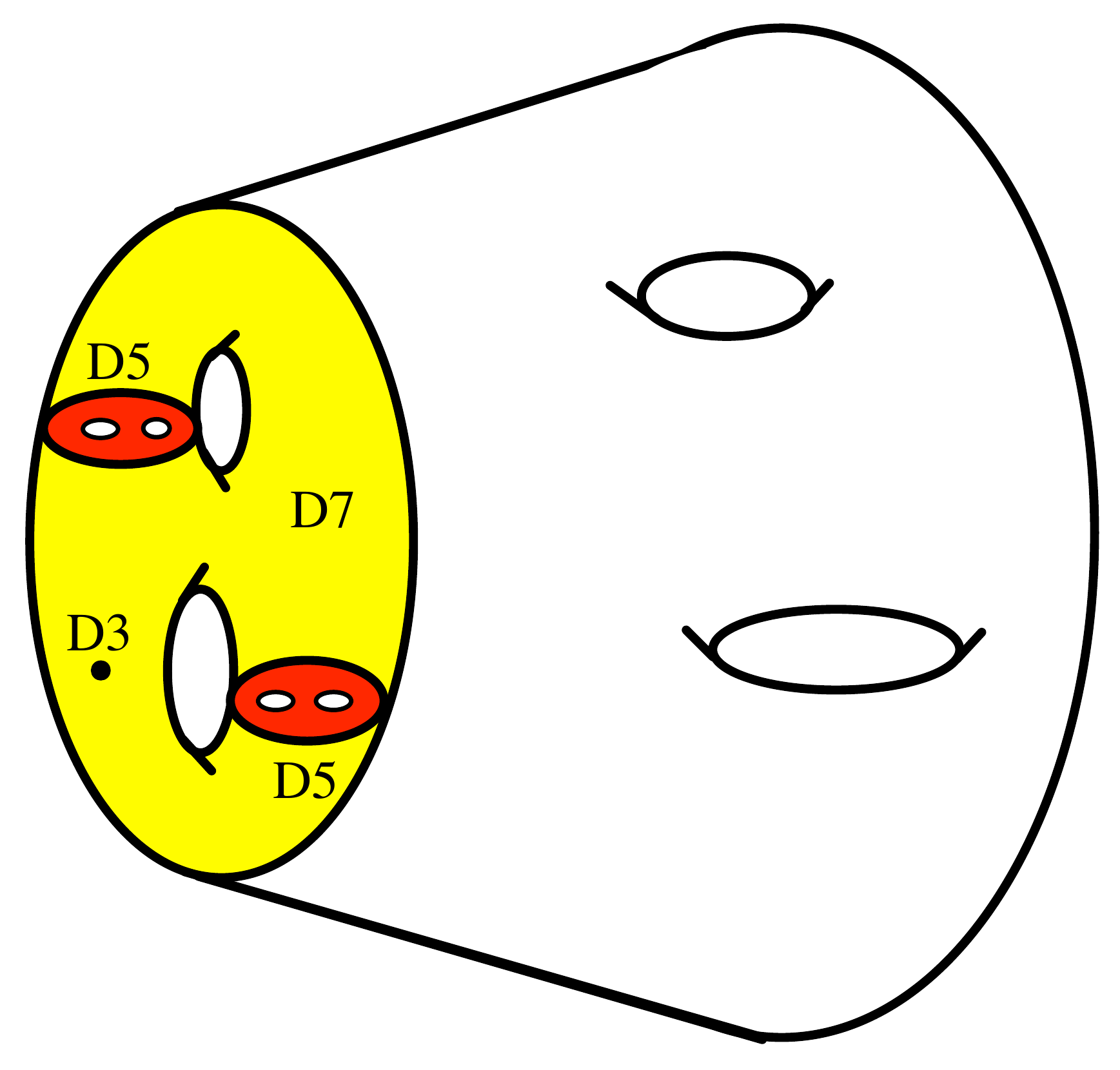,scale=0.27} \vspace{-6mm}
\end{center}
\noindent
\caption{D(p+3)-branes may wrap $p$-cycles within the base $X$ of the Calabi-Yau singularity.}
%\label{default}
\vspace{-6mm}
\end{figure}

\medskip

\subsection{D-branes and fractional branes}

The D-branes are charged with respect to the Ramond-Ramond (RR) fields
\cite{Polchinski:1995mt}.
In flat space the interaction between the Dp-brane and the RR
fields is summarized by the Chern-Simons terms in the action
\cite{Polchinski:1996na}
\be\lb{flatCS}
S_{CS}=\mu_p \int e^{F-B}\sum_n C^{(n)}.
\ee
Here the integral goes over the world volume of the brane, the
index $n$ runs over even integers for type IIB and over odd integers
for the type IIA. The fields $C^{(n)}$ are the RR n-form
fields. $F$ is the gauge field on the world volume of the brane and
$B$ is the restriction of the $B$-field to the brane.

In a more general settings, one may consider a stack of $N$ D-branes
wrapping some cycle $M$.
It is convenient to introduce an $N$-dimensional vector bundle $V$
over $M$.
The coordinates of the vectors $v=(v_1,\ldots, v_N)$ in $V$ are labeled by the
the Chan-Paton index $i=1\ldots N$.
The gauge field $F\in SU(N)$ living on the world volume of the stack
of branes is interpreted as the curvature form for $V$.
(It is a two-form on the tangent bundle $TM$ that takes
values in the operators acting on $V$.)

A priori, a D-brane can be placed on any subspace of the 10-dimensional
space-time on which the string theory lives.
But in general, such a configuration will not be stable.
The stable objects are the BPS branes.
The BPS branes have a minimum energy in a given topological class
and preserve some supersymmetry.

One of the main questions is to find the set of BPS branes for a
given CY geometry.
If one changes the geometry or moves a brane around the manifold,
then the BPS brane can become non stable and decay to a combination
of new BPS branes.
For example, a D3 brane is stable at a smooth point in CY but it
decays to a combination of so-called fractional branes near a singularity.\footnote{
The terminology `fractional brane' is motivated as follows.
Near the $\C^3/Z_3$ singularity, D-branes must form a representation
of $Z_3$, e.g. the $Z_3$ can act by interchanging the branes placed
at 3 image points.
The BPS branes at the singularity may be called fractional because a
D3-brane splits into three branes, each carrying a 1/3 of the
D3-brane mass.}
In terms of the world-sheet CFT, that describes the string propagation on the singularity,
they are in one-to-one correspondence with the allowed conformally
invariant open string boundary conditions. Alternatively, by extrapolating to a large volume
perspective, fractional branes may be represented in geometrical language as particular well-chosen
collections of sheaves,  supported on corresponding
submanifolds within the local Calabi-Yau singularity.
A geometric way of classifying the space of possible fractional branes at a
Calabi-Yau singularity is in terms
of the "hidden" cycles that the branes can wrap.
The hidden cycles can be seen by resolving the singularity.
%The $\C^3/Z_3$ singularity is resolved by putting a $\PP^2$ that
%has three independent cycles: a 4-cycle, a 2-cycle and a 0-cycle.
The fractional branes can be defined as appropriate stable bound states of branes
wrapping these cycles.
%Note, that these bound states are BPS only for the small volume of
%$\PP^2$.
%For the large volume only the pure branes wrapping a 4-cycle, a
%2-cycle, or a 0-cycle are BPS, but any combination of them brakes
%the supersymmetry and is not BPS.
These bound states, in turn, can be thought of as a D-brane with some
quantized magnetic flux $F$ supported on cycles wrapped by  the worldvolume
of the maximal dimension brane.
The form $F$ is the Poincare dual form to the cycles on which the
sub-branes are wrapped.\footnote{Let us recall the definition of the Poincare dual form.
Let $M$ be a compact complex manifold and $A$ be a cycle inside $M$.
The form $w_A$ is said to be the Poincare dual to $A$ inside $M$
if for any form $\al$
\bea
\int_A \al=\int_M w_A\wedge \al
\eea
For example, the hyperplane class $H$ in $\PP^2$ is the two-form
Poincare dual to $\PP^1\subset \PP^2$.}

Topologically non trivial configurations of $F$ are interpreted as
lower dimensional branes bound to the stack of $N$ D-branes.
The charges of these lower dimensional branes can be determined from
the interaction with the RR fields.
Let us introduce the notation for the charge vector
\be\lb{chvec}
{\cal Q}=\ch(V)\sqrt{\frac{\hat A(T)}{\hat A(N)}}
\ee
%\be\lb{CSE}
%L_{CS}=\int_M \Tr(e^{F})\sqrt{\frac{\hat A(T)}{\hat A(N)}} \sum_n C^{(n)}
%\ee
where
\be
\ch(V) = \Tr(e^F)
\ee
is the Chern character of the vector bundle $V$.
The first three characters are the rank, the first Chern class and
the "instanton number"
\be
\barr{lll}
\ch_0(V)=\rk(V)=N\\[2mm]
\ch_1(V)=c_1(V)=\Tr(F)\\[2mm]
\ch_2(V)=\frac{1}{2}\Tr(F\wedge F)
\earr
\ee
The term in the square root in (\ref{chvec})
is related to the curvature of the
D-brane. It is necessary for the cancelation of gravitational
anomalies \cite{CheungYin}.
The charge vector can be expanded in terms of the cohomology
classes.
In the case of del Pezzo surfaces there are several 2-cycles $\al_A$,
consequently the D2-brane charges will come with an
index $A$, labeling the cycles that the brane wraps.
In terms of their Poincare dual forms the charge vector is expanded as
\ba\lb{charge}
{\cal Q} =Q_7 + %\sum_A
Q_5^A\om_A + Q_3 H\wedge H
\ea
where $H$ is the hyperplane class on $\PP^2$.
The first number represents the D7-brane charge (as always, we include
the four non compact dimensions in the world volume of the branes),
the second element is the Poincare dual class to the cycle that the
D5-brane wraps, and the last number is the D3-brane charge.
The linear  Chern-Simons coupling takes the form
\bea
L_{CS}=Q_7\int_{\PP^2} C^{(8)}+%\sum_A
Q^A_5\int_{\al_A} \!\! C^{(6)}
+Q_3 \:C^{(4)}
\eea
in accord with the identifications of $Q_p$ with the wrapped $Dp$-brane charge.

\newcommand{\is}{\! &\! = \! &\!}

A D3-brane placed at the tip of the cone can split into fractional branes.
This splitting is possible since it doesn't violate the conservation of the
charges.
Whether the splitting will actually happen is a more difficult
question.
The answer to this question involves some knowledge about the masses
of the branes.
The masses of the BPS objects are proportional to the absolute value of their
central charges.

%The situation here is similar to \cite{Seiberg:1994rs} where the
%mass of a state is determined by the charges $q_e$ and $q_m$.
%If vacuum values of the background fields are $A$ and $A_D$, then the
%mass is given by the absolute value of the central charge
%$m=|Z|$, where
%\be\lb{SWcc}
%Z(q_e,q_m)=q_e A+q_m A_D
%\ee
The Born-Infeld action for the D-branes in the BPS limit reduces to
the absolute value of the central charge,
which in the large volume limit has the form \cite{Douglas:2000qw}
\be\lb{gencc}
Z(V)=\int\! e^{-B-iJ}\ch(V)\sqrt{\frac{\hat A(T)}{\hat A(N)}}
\ee
here $B+iJ$ is the complexified Kahler form which is determined by the
background geometry. % and plays the role of $A$ and $A_D$.
The central charge then takes the form
\be\lb{branecc}
Z(V)=\Pi_0Q_7+\Pi^A_2Q^A_5+\Pi_4Q_3
\ee
%It is interesting to note the similarity between the terms
%in (\ref{SWcc}) and (\ref{branecc}).
The charges $(Q_7,Q_5^A,Q_3)$ can be found by expanding the
charge vector (\ref{chvec}) in terms of the forms as in
(\ref{charge}).
In the large volume limit the periods are found by expansion of
$e^{-B-iJ}$
\ba
\nonumber
\Pi_0&=&1\\
\lb{periods}
\Pi_2&=&-\int_{\PP^1}(B+iJ)\\
\nonumber
\Pi_4&=&\frac{1}{2}\int_{\PP^2}(B+iJ)\wedge (B+iJ).
\ea
These expressions for the periods in terms of the background
fields $B$ and $J$ receive non perturbative corrections
away from the large volume limit.

The central charge is an important characteristic of the D-branes, as it
tells what supersymmetry is preserved (broken) by the branes.
The couplings and the FI terms of the gauge theory in dimensional
reduction also depend on the central charges.

\bigskip

{\small
\noindent
{\bf  Example 1: Branes on a CY cone over $\PP^2$.}\\
In this example we find the charge vectors for the fractional branes
on the CY cone over $\PP^2$, this cone is equivalent to a
$\C^3/Z_3$ singularity \cite{Douglas:2000qw}.
The charge of the brane at the tip of the cone can be expressed in
terms of the cohomologies of $\PP^2$
\be
{\cal Q}=Q_7 + Q_5 H + Q_3 H\wedge H.
\ee
where $H$ is the hyperplane class of $\PP^2$.
If we have $N$ D7-branes wrapping the $\PP^2$ in the blowup of
$\C^3/\Z_3$, then depending on the fluxes of the gauge field $F$,
i.e. depending on the character of the bundle $V$,
the components of the charge vector are \cite{Diaconescu:1999dt}
\bea
\begin{array}{lllll}
Q_7&=&\rk(V)\:=\:N\\[2mm]
Q_5&=&\int_{\PP^1}c_1(V)\\[2mm]
Q_3&=&\int_{\PP^2}\ch_2(V)+\frac{1}{8}\rk(V)
\end{array}
\eea
The last term for $Q_3$ comes from
\cite{CheungYin}\cite{Diaconescu:1999dt}
\bea
\sqrt{\frac{\hat A(T)}{\hat A(N)}}
=1+\frac{\chi(\PP^2)}{24}H\wedge H
\eea
where the Euler character $\chi(\PP^2)=3$.
This term can be
interpreted as the D3 brane charge induced by the curvature of
the branes.

The calculation of the periods in the small volume limit is a non
trivial problem since they receive non perturbative corrections.
The periods and the central charges for the branes on $\C^3/Z_3$
singularity can be found e.g. in
\cite{Diaconescu:1999dt}.
}

\medskip
\medskip

{\small
\noindent
{\bf Example 2: D3 at the tip of the cone over $\PP^1\times \PP^1$.}

In this example we find the central charges of the fractional branes
on the cone over $S=\PP^1\times \PP^1$.
We denote by $H_1$ and $H_2$ the 2-cycles Poincar\'e dual to the two
$\PP^1$'s.  They have intersections
\bea
H_1^2 = H_2^2 = 0,  \qquad H_1 \cdot H_2 = 1\, .
\eea
The canonical class is
\bea
K(S)=-2(H_1+H_2).
\eea
Line bundles over the base $S$ are of the general form
\bea
{\cal O}(n,m) = {\cal O}(n H_1 + m H_2) \, .
\eea
In other words, if we choose coordinates
on the first $\PP^1$ as $z_\alpha$, $\alpha = 1, 2$ and
on the second $\PP^1$ as $w_\beta$, $\beta = 1, 2$, then
sections of $H_0( {\cal O}(n, m))$ are
polynomials $P (z , w)$ of total degree $n$ in $z$ and total degree
$m$ in $w$ (assuming $n, m\geq 0$).

The basis of fractional branes on $S$ is given by appropriate sheaves. In our
example the
sheaves are particularly simple: they are given by the following set of line bundles
\ba
\label{basis}
{\mathsf F_1} =  {\cal O}(0)\, , \qquad & & {\mathsf F_3}=   {\cal O}(H_2)\, , \\[2mm]
{\mathsf F_2} = {\cal O}(H_1)\, ,\quad & &  {\mathsf F_4} = {\cal O}(H_1+H_2)\, , \nonumber
\ea
and carry the following set of  charge vectors
\ba
\lb{p1p1branes}
{\mathsf F_1}%Q(\OO(0))
=(1,0,0) \, , \qquad &&
{\mathsf F_3} = (1,H_2,0)\, ,\\[2mm]
{\mathsf F_2} =(1,H_1,0)\, ,\quad & &
{\mathsf F_4} %Q(\OO(E_1+E_2))
=(1,H_1+H_2,1) \, . \nonumber
\ea
Each charge vector indicates a corresponding bound state of wrapped D-branes.

Charge conservation lets
the D3-brane split into four fractional branes via
\be
\label{deco}
(0,0,1) = {\mathsf F_1}-{\mathsf F_2}-{\mathsf F_3}+ {\mathsf F_4}
\ee
Let us find the central charges of these fractional branes and
verify that the splitting is possible from the point of view of the
masses.
For the cone over $\PP^1\times \PP^1$
the central charge is
\be\lb{cencha}
Z=\pm\int_S e^{F-B}
\ee
where the plus sign is for the branes and minus is for the antibranes.

If we blow up one of the $\PP^1$'s, then the geometry near the
second $\PP^1$ will be $\C\times\C^2/Z_2$.
The $\PP^1$ in the blow up of the singularity is the second $\PP^1$
in $S$.
It is known \cite{Aspin-orbi} that as the $\PP^1$ shrinks to form the $Z_2$
singularity the value of the $B$ field period is $1/2$, i.e. at the
$Z_2$ orbifold point
$\int_{\PP^1}(B+iJ)=\frac{1}{2}$.
We will assume that changing the size of the first $\PP^1$ doesn't
affect the periods over the second one,
then in the limit $\vol(S)=0$ the value of the $B$ field is
\be
B=\frac{1}{2}(H_1+H_2).
\ee
As an example of the calculation, let us find the central charge for
the antibrane ${\bf \bar F_2}=-(1,H_1,0)=-e^{H_1}$
\bea
Z({\bf \bar F_2})&=&-\int_{S}e^{H_1-B} = -\int_S e^{{1\over 2} (H_1-H_2)} \\
%&=&\frac{1}{2}\int_{\PP^1\times\PP^1} (E_1-B)\wedge (E_1-B)\\
&=&\frac{1}{4}\int_{\PP^1\times\PP^1} H_1\wedge H_2\; =\, \frac{1}{4}
\eea
It's easy to check that the central charges of the other three
configurations are also equal to $\frac{1}{4}$.
The sum of the masses is equal to $1$ which is the mass
of the D3-brane.
The phases of the central charges are the
same, i.e. the corresponding fractional branes break/preserve the same
supersymmetry generators.}

\bigskip

\section{Quiver gauge theories}

%So far we have considered:\\
%1. Conditions for the background to be CY\\
%2. Characterization of the branes in terms of their RR charges\\
%3. The masses and the central charge of the branes.
%Now we are going to consider systems of branes.

\def\Tr{{\rm Tr}}
\def\Hom{{{\rm{Hom}}}}

\def\tr{{{\rm{tr}}}}
\def\rchi{{\hbox{\raise1.5pt\hbox{$\chi$}}}}
\def\Aut{{{\rm{Aut}}}}
\def\isom{\cong}

\newcommand{\GG}{G}

A collection of fractional branes gives rise
to a quiver gauge theory.
In the absence of orientifold planes
the quiver is a graph with oriented edges.
Every fractional brane corresponds to a vertex
in the quiver. If there are $N$ fractional branes of the same sort,
then they correspond to the $U(N)$ gauge group.
An edge in the graph starting on $U(N_1)$ and ending on $U(N_2)$
corresponds to the bifundamental field $\Phi\in(\bar N_1, N_2)$, where
$\bar N_1$ denotes the antifundamental representation of $U(N_1)$
and $N_2$ is the fundamental represenation of $U(N_2)$.
Every edge corresponds to a massless mode of open strings
stretching between fractional branes. The lightest open string modes are massless
whenever the two branes intersect with each other.
The orientation of the open string is translated in the orientation
of the edge.
Note that there can be several edges between two vertices, also an
edge can begin and end on the same vertex, in this case the field is
in adjoint representation of the corresponding gauge group.

If there are orientifold planes, then some of the open strings
become unoriented.
The corresponding gauge groups are $SO(N)$ or $Sp(N)$ and the fields
are in real representations of these gauge groups.

The orientation of the edge also corresponds to the chirality of the
bifundamental field.
At every vertex, the number of incoming and outgoing edges is
the same.
This property ensures that there are no cubic anomalies, i.e.
anomalies with three $SU(N)$ currents.
But if there is a net number of chiral fields between two
vertices, then the $U(1)$ parts of the corresponding $U(N)$ gauge groups
have mixed anomalies.
Note that some combinations of the anomalous $U(1)$'s can be non
anomalous.\footnote{For the cones over del Pezzo surfaces the
combination of $U(1)$'s is not anomalous if the corresponding sum of
fractional branes has no D7-brane charge and the class of the
D5-brane doesn't intersect the canonical class of the del Pezzo.
In short the argument goes as follows.
The divisor for the normal bundle over $X$ is the canonical
class. Thus the normal bundle is non trivial over $X$
and over all cycles that intersect the canonical class $K$.
If a cycle doesn't intersect $K$, then the normal bundle over this
cycle is trivial:
%, i.e. such cycle generically
it doesn't intersect with
any other cycle. % within the cone over dell Pezzo (in fact it could
%intersect only with the 4-cycle, i.e. the dell Pezzo itself).
If the two fractional branes don't intersect, then there is no
chiral matter between them.
Hence the corresponding $U(1)$ is non anomalous.}

Before we go to some practical details on finding the quiver gauge
theory, let us mention the question of stability \cite{Douglas:2000ah}.
The conservation of the mass and the charge is a necessary but not
a sufficient condition for a splitting of a brane into fractional branes
to exist.
The problem is that some fractional branes may further split into
sub-branes or may form a new bound state.
The collection of fractional branes should be stable against further
reductions.
For a mathematical description of various stability condition see
for example \cite{Douglas:2000gi}\cite{Douglas:2000ah}.
From the point of view of the corresponding quiver gauge theory
stability means that there
are no adjoint fields and that for any two vertices all the edges
between them have the same orientation (if there are any).
The last condition is, in fact, more strict: there should exist an
order of the fractional branes (let the order be from left to right)
such that the orientation of the edge is from the left fractional
brane to the right one.
The collection of fractional branes that satisfy the stability conditions
is called exceptional.

\bigskip

{\small
\noindent
{\bf Example 1: D3 near an orbifold singularity}

A useful illustration of how quiver gauge theories arise is provided by the example of a D3-brane
near a general orbifold singularity \cite{Douglas:1996sw}\cite{Lawrence:1998ja}.
Let $\GG$ be some finite group  of order $|\GG|$,  that acts on ${\bf C}^3$.
$\GG$ can be abelian or non-abelian.
%We wish to consider the world-volumetheory of a D3-brane near the orbifold singularity.
In case $\GG$ is a sub-group of $SU(3)$,
the world-volume theory is ${\cal N}=1$ supersymmetric.
To find states invariant under the orbifold projection, we have to consider the D3-brane
and all of its images, making a total of $|\GG|$  D3-branes. From now on let us denote
\be
N = |G|
\ee
Before performing the orbifold
projection, the world-volume theory on the $N$ D-branes is a $U(N)$ gauge theory
with a vector multiplet $V$ and three chiral multiplets $\Phi_i$, that parametrize the
transverse positions of the D3-branes along  ${\bf C}^3$. All fields are
 $N\times N$ matrices.
Projecting onto $\GG$  invariant states amounts to imposing the conditions
\ba
\label{vproj}
R_{reg} V R_{reg}^{-1}, = \, V \\[1mm]
(R_3)_{ij} R_{reg} \Phi^j R_{reg}^{-1}\is \Phi^i \label{phiproj}
\nonumber
\ea
where $R_{reg}$ is the $N\times N$ regular representation of $\GG$ acting on the
Chan-Paton index, and $R_3$ is the 3-d defining representation.
The regular  representation is defined as the group $\GG$ acting on itself.
\def\Tr{{\rm Tr}}
\def\Hom{{{\rm{Hom}}}}

\def\tr{{{\rm{tr}}}}
\def\rchi{{\hbox{\raise1.5pt\hbox{$\chi$}}}}
\def\Aut{{{\rm{Aut}}}}
\def\isom{\cong}

%The group multiplication on $G$ induces a multiplication
%rule on ${\bf C}[G]$ via
%\be
%x \cdot y = \sum_{\gamma_1, \gamma_2 \in G^2} x(\gamma_1) y(\gamma_2) \gamma_1 \gamma_2
%\ee

The regular representation $R_{reg}$ is not irreducible; instead it decomposes into irreducible representations as
\be
\label{decomp}
R_{reg}=\bigoplus_{a=1}^r n_a R^a
\ee
where $r$ denotes the total number of irreducible represenations and
\be
n_a={\rm dim} R^a.
\ee
In other words, each irreducible representation $R^a$ occurs $n_a$
times in the regular representation.  In explicit matrix notation, we have
\be R_{reg}=  \mbox{\footnotesize $
\left(\begin{array}{cccc}{R^1 \otimes 1_{n_1}} & {0} & ... & 0 \\
0 & R^2 \otimes 1_{n_2} & ... & 0 \\
\vdots& \vdots & \ddots & \vdots \\
0 & 0 & ... & R^r \otimes 1_{n_r}  \end{array} \right) $}
\ee
where $R^a \otimes 1_{n_a}$ is the $n_a^2 \times n_a^2$ matrix
\be
R^a \otimes 1_{n_a} =  \mbox{\footnotesize $
 \left( \begin{array}{cccc} R^a & 0 & ... & 0 \\
0 & R^a & ... & 0 \\
\vdots& \vdots & \ddots & \vdots \\
0 & 0 & ... & R^a  \end{array}\right) $}
\ee
From this form of $R_{reg}$ we read off that the (\ref{phiproj}) breaks the
$U(N)$ gauge symmetry to
\be
\label{ggroup}
\prod_{a=1}^r U(n_a)
\ee
Translated into geometric language,
we conclude that a D3-brane near an orbifold singularity splits up into fractional branes $F_a$,
where $a$ labels an irreducible representation $R_a$, and that each fractional brane
occurs with multiplicity $n_a = {\rm dim} R_a$.
The worldvolume theory of each fractional brane $F_a$ contains a vector multiplet $V_a$
which in particular is an $n_a\times n_a$ matrix. This result is a reflection of the decomposition
of the group algebra as a direct sum of $n_a \times n_a$ matrices
\be
\label{decom} {\bf C}[G] \isom \bigoplus_{a=1}^r {\rm
Mat}( n_a)\;,
\ee
which for us states that the vector multiplets $V_a$, when all combined together, can be thought
of as an element  of ${\bf C}[G]$. %We will return to this identification momentarily.
%as a simple illustration of a deep correspondence known as open/closed string duality.

From the condition (\ref{phiproj}), we learn that we can obtain the number of chiral fields
$n^3_{ab}$ between two fractional branes $F_a$ and $F_b$, transforming in the $(n_a, \overline{n}_b)$ bi-fundamental representation, by decomposing the
product of the defining and each irreducible representation into irreducible
representations in the following way:
\be
R_3 \otimes R^a = \bigoplus_{b=1}^r n^3_{ab} R^b.
\label{couplings}
\ee
Using that the multiplication of group characters reflects the representation algebra of the group,
we can compute these coefficients $n^3_{ab}$ as
\be
\label{number}
n^3_{ab}=\frac{1}{|G|}\sum_{g \in G} \chi^3(g)\chi^a(g) \chi^b(g)^*,
\ee
where we used the orthogonality condition of group characters
\bea
\frac{1}{|G|}\sum_{g \in G} \chi^a(g) \chi^b(g)^*=\delta_{ab}.
\eea
Eqns (\ref{ggroup}) and (\ref{number}) provide the complete quiver data of the D3-brane gauge theory.

\bigskip

\smallskip

\noindent
{\bf Example 2: D3 on a toric $dP_n$ singularity}

As a warm-up, we first summarize the D3-brane gauge theory on a $dP_n$ singularity
with $n\leq 3$.
These all admit an elegant and useful diagrammatic description in terms of $(p,q)$-webs
\cite{Feng:2004uq}.

A toric manifold of complex dimension 2 can be characterized as a torus fibration over the
complex plane. The torus admits two $U(1)$ actions, with generators $T_1$ and $T_2$.  The $U(1)$'s
may have fixed points at some special locus within the complex plane, at which
the fibration becomes degenerate. This locus can be drawn as a so-called $(p,q)$
web, which is a trivalent graph of lines and vertices. Each line has  an associated charge vector
\be
v_i = \left(\!\! \begin{array}{c} p_i \\ q_i \end{array} \!\! \right).
\ee
Here $p_i$ and $q_i$ two relatively co-prime integers,  that parametrize the linear combination
$p_i \,T_1 + q_i \, T_2$ of the two $U(1)$  generators that degenerates at the location of the
line. The lines meet at trivalent vertices, which are special points where both
$U(1)$ actions degenerate. The charge vectors of three lines that meet at any given
vertex add up to zero.

The charge vector $v_i$ of each line also indicates its direction. This identification makes use
of the fact that the del Pezzo surface has a symplectic form, represented by the K\"ahler class,
relative to which the complex plane, the base, can be thought of as the space of coordinates and the
torus fiber as the space of momenta. We can therefore make a natural identification
between directions within the complex plane and directions along the $T^2$ fiber.

The $(p,q)$-web spans the whole complex plane, and therefore has lines that extend to infinity,
called external lines, as well as internal lines that are compact intervals. The internal
lines are the base of a circle fibration over an interval, and thus represent compact two-cycles.
Similarly, a face of the (p,q)-web,  a compact region of the complex
plane bounded by compact lines, represents a compact 4-cycle.

The quiver gauge theory data are encoded in the $(p,q)$-web as follows.
Nodes on the quiver correspond to fractional branes, which are the possible bound states
of wrapped D3, D5 and D7 branes. Each node represents a $U(1)$ gauge factor (or $U(N)$
in the case one considers $N$ D3-brane probes).
The number of independent fractional branes equals
the dimension of the homology of the surface, that is, the number of 2-cycles plus 2. For
$dP_n$ this adds up to $n+3$. This equals the number of external lines
of the corresponding $(p,q)$-web. Indeed, it turns out
that there is a one-to-one correspondence between fractional branes $F_i$ on $dP_n$
and  the  vertices of its $(p,q)$-web that are connected to external lines. To each $F_i$ we can
thus associate a charge vector $v_i = (p_i,q_i)$,  equal to the direction of the corresponding
external line. The intersection pairing between two fractional branes, which counts the
number of bi-fundamentals between them, is then expressed as follows
\be
\label{pqpair}
\#(v_i,v_j) = p_i q_j - q_i p_j
\ee

The quiver gauge theories associated with the toric del Pezzo surfaces are somewhat
trivial, since they only contain $U(1)$ gauge factors for the case of a single D3-brane.
Nonetheless, they provide some useful general guidelines for how to translate gauge
theory data into geometric data, and vice versa. }

\bigskip

\subsection{D3-brane on a $dP_n$ singularity}
%Suppose we know an exceptional collection of fractional branes
%that corresponds to splitting of a D3-brane at the singularity.
%Then it is relatively easy to find the corresponding gauge theory.

Consider a D3-brane placed at the tip of the complex cone over
a dell Pezzo surface.
The D3-brane will split in an exceptional collection of fractional
branes.
Let $\FF_i$ denote the charge vector of the $i$th fractional brane
in the collection, $i=1\ldots m$.
Recall that a fractional brane correponds to a vector bundle $V_i$
such that the charge vector (\ref{chvec}) for the interaction with
the RR fields is
\be
{\cal Q}_i=(\rk(V_i), \;c_1(V_i), \;\ch_2(V_i) +\rk(V_i) c
)
\ee
where $c$ is a certain constant induced by the curvature of del Pezzo.
The decomposition of the D3-brane charge introduces a set of multiplicities
$N_i$ via
\be
\label{decos}
(0,0,1)=\sum_{i=1}^m N_i\; {\cal Q}_i.
\ee
%where $N_i$ is the multiplicity of $\FF_i$ in the decomposition.
Note, that the total $D7$-brane charge is zero $\sum_i N_i\; \rk(V_i)=0$.
This means that some of the $N_i$'s are negative: this corresponds to
taking $|N_i|$ antibranes $\bar\FF_i$.
Since there are $m$ fractional branes in the decomposition, the
quiver gauge theory consists of a product of $m$ gauge groups
\be
G = \prod_{i=1}^m U(N_i).
\ee

The next step is to find the matter fields.
Consider two fractional branes $\FF_i$ and $\FF_j$.
For the exceptional collection, if there are some fields $\Phi_{ij}$
in the representation $(\bar N_i, \;N_j)$, then there are no fields
in the opposite representation $(\bar N_j, \;N_i)$ and the total
number of chiral fields between $\FF_i$ and $\FF_j$ is given by the
intersection of these fractional branes.

In order to define the intersection number let us introduce the
notation for the intersection of the 2-cycle that the $\FF_i$
wraps with the canonical class of the del Pezzo surface
\be
\deg(\FF_i)=c_1(V_i)\cdot K.
\ee
Then the matrix of intersections between the fractional branes is given by
(here $\rk(\FF_i)=\rk(V_i)$)
\be
\label{int}
N_{ij}=\rk(\FF_j)\deg(\FF_i)-\rk(\FF_i)\deg(\FF_j).
\ee
The absolute value $|N_{ij}|$ is the number of edges between $\FF_i$
and $\FF_j$.
The direction of the edges is determined by the sign of $N_{ij}$.

The geometrical motivation of the formula (\ref{int})
% can be found in
%\cite{Verlinde:2005jr}, \cite{Diaconescu:1999dt}
%\cite{Douglas:2000ah}.
%Intuitively this formula can be motivated
is as follows.
Let us find the intersection of $\FF_i$ with $\FF_j$ by deforming
the cycles in $\FF_i$ along the normal directions to the del Pezzo surface $X$.
The surface $X$ intersects itself via the canonical class. Consequently,
the D7-brane component of $\FF_i$ intersects $\FF_j$ via its component along the
canonical class. Since this component is wrapped ${\rm deg}({\mathsf F}_j)$ times,
and the D7-brane charge of ${\mathsf F_i}$ equals ${\rm rk}({\mathsf F_j})$, the intersection
number receives a contribution equal to the product ${\rm deg}({\mathsf F_j}) \, {\rm rk}({\mathsf F_i})$.
The same logic works for the intersection of D5 component of
$\FF_i$ with D7 component of $\FF_j$.

Using the langauge of algebraic geometry, one can give a
more invariant way of finding the chiral matter in the collections
of type IIB branes.
In general, let $V_1$ and $V_2$ be two vector bundles associated to
two fractional branes $\FF_1$ and $\FF_2$.
Open strings from $\FF_1$ to $\FF_2$ have one Chan-Paton index in
$V_1$ and the other in $V_2$. They correspond to homomorphisms
from $V_1$ to $V_2$, i.e. a chiral matter field associated with edge
between the quiver nodes $\FF_1$ and
$\FF_2$ is an element of $\Hom(V_1,V_2)$.
More generally, the fields that live on the intersections of
two fractional branes $\FF_1$ and $\FF_2$ are
represented by the so-called extension groups $\Ext^p(\FF_1,\FF_2)$.
The zeroth extension group is spanned by the homomorphisms
\ba
\Ext^0(\FF_1,\FF_2)=\Hom(\FF_1,\FF_2).
\ea
%The fields for the higher extension groups are interpreted as
For the exceptional collections of branes, all higher extension
groups are trivial.

 \begin{figure}[th]
\begin{center}
        %\resizebox{\textwidth}{!}{
           \epsfig{figure=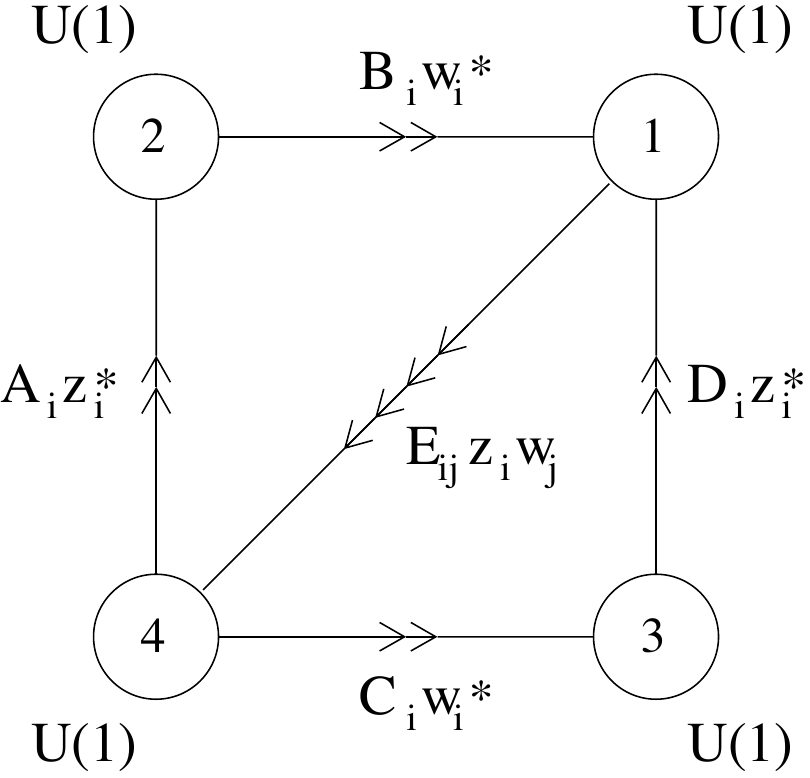,scale=0.55}
           % \scalebox{0.7}{
         %\includegraphics[scale=0.55]{pdf-files//p1p1}
\end{center}
\vspace{-5mm}
\caption{Quiver theory of the cone over $\PP^1\times\PP^1$.}
\label{p1p1fig}
\vspace{-5mm}
\end{figure}

\bigskip

{\small

\noindent
{\bf Example: D3 at the tip of the cone over
$\PP^1\times\PP^1$.}

Let us return to our example of a D3-brane at the tip of a cone over
$\PP^1\times\PP^1$.
The basis of fractional branes is specified via the charge vectors given in (\ref{p1p1branes}).
The degree is defined by the intersection of D5-brane charge
with the canonical class, $K=-2(H_1+H_2)$. So we have\footnote{Here
we have flipped the sign of ${\mathsf F}_2$ and
${\mathsf F_3}$, since these occur with negative multiplicity in (\ref{deco}).}
\be
\barr{lll}
\rk(\FF_1)=1\;\;\;&\deg(\FF_1)=0\\[2mm]
\rk(\FF_2)=-1\;\;\;&\deg(\FF_2)=2\\[2mm]
\rk(\FF_3)=-1\;\;\;&\deg(\FF_3)=2\\[2mm]
\rk(\FF_4)=1\;\;\;&\deg(\FF_4)=-4
\earr
\ee
Every fractional brane corresponds to a gauge group.
Since in the decomposition (\ref{deco}) of a single D3 brane, all
fractional branes    occur with multiplicity $\pm 1$, the
gauge group of the gauge theory on the D3-brane is $U(1)^4$.
The number of chiral fields between two fractional branes is
given by their oriented intersection number.
Via the general formula (\ref{int}), we find
\bea
\#(\FF_i,\FF_j)= %\rk(\FF_2)\deg(\FF_1)-\rk(\FF_1)\deg(\FF_2)
\left(\! \begin{array}{cccc}
\ 0 & -2 & -2 & \ 4  \\[1mm]
\ 2 &\ 0 &\ 0 & - 2  \\[1mm]
\ 2 &  0 & \ 0 & -2 \\[1mm]
-4 & \ 2 & \ 2 & \ 0  \\
\end{array}\!
\right)
\eea
The resulting quiver gauge theory is given in figure \ref{p1p1fig}.

The number of chiral matter fields between the fractional branes is equal to the
dimension of the corresponding space of homomorphisms. Recall
the definitions (\ref{basis}) of the basis of fractional branes, and that,
if we choose coordinates
on the first $\PP^1$ as $z_\alpha$, $\alpha = 1, 2$ and
on the second $\PP^1$ as $w_\beta$, $\beta = 1, 2$,
sections of $H_0( {\cal O}(n, m))$ are
polynomials $P (z , w)$ of total degree $n$ in $z$ and total degree
$m$ in $w$ (assuming $n, m\geq 0$).
Let us denote the operators dual to $z_1$, $z_2$ by $z_1^*$,
$z_2^*$, etc.
That is, $z_1^*$ is the linear operator that maps $z_1$ to 1, and all other coordinates
to 0, etc. Then
\bea
%\Hom(\FF_3,\FF_1) %&=& %\Hom(\OO(0,0),\OO(1,0))^*\\
%&=&C_1 z_1^*+C_2 z_2^*\\[1mm]
\Hom(\FF_2,\FF_1)%&=&\Hom(\OO(0,0),\OO(0,1))^*\\
&=&B_1 w_1^*+B_2 w_2^*\\[1mm]
\Hom(\FF_4,\FF_2)%&=&\Hom(\OO(1,1),\OO(0,1))^*\\
&=&A_1 z_1^*+A_2 z_2^*\\[1mm]
\Hom(\FF_3,\FF_1)%&=&\Hom(\OO(0,0),\OO(1,0))^*\\
&=&D_1 z_1^*+D_2 z_2^*\\[1mm]
\Hom(\FF_4,\FF_3)%&=&\Hom(\OO(1,1),\OO(1,0))^*\\
&=&C_1 w_1^*+C_2 w_2^*\\[1mm]
\Hom(\FF_1,\FF_4)%&=&\Hom(\OO(0,0),\OO(1,1))\\
&=&\mbox{$\sum_{i,j}$} E_{ij} z_iw_j
\eea
%where the sum is over $i,j=1,2$.
We will use these expressions later to
compute the superpotential of the quiver gauge theory of Fig 2.

Note, that the only non anomalous combination of the $U(1)$ groups is
the difference between the $U(1)$ at the nodes 3 and 2.
The fields $A$ and $D$ have the charge $+1$, the fields $B$ and $C$
have the charge $-1$, and the field $E$ is neutral under this $U(1)$.
The corresponding combination of the fractional branes is
\be
\FF_\al=\FF_3-\FF_2=(0,E_2-E_1,0).
\ee
This combination is the only combination
of fractional branes that has $\rk(\FF_\al)=0$ and
$\deg(\FF_\al)=0$.
Consequently it has no chiral intersection with any other fractional
brane in the collection.
This is an example of a more general statement, that the
combination of fractional branes with no D7-brane charge and with the D5-brane
charge that has no intersection with the canonical class corresponds
to a non anomalous combination of the $U(1)$ gauge groups.
}

\bigskip

\section{Geometric Identification of Couplings}

Consider a D-brane placed at a
singularity of a compact Calabi-Yau manifold. We can take a formal low energy limit,
in which  the distances that can be probed by the open strings are much
smaller than the size of CY. In this decoupling limit, we may focus our
attention to the local region of the singularity, which for simplicity
we can take to be non-compact. In this limit the kinetic terms of the closed
string modes propagating on the full CY become non normalizable, and the
corresponding fields enter in the action as true non dynamical
parameters.
%(these VEVs can depend on the dynamics of the other fields through the RG flows etc.).
If the kinetic term of the closed string mode is localized near the
singularity, then the corresponding field remains dynamical in the
effective theory.

\bigskip

\subsection{Superpotential}

In our discussion thus far, we have concentrated our attention on the topological properties of D-branes on Calabi-Yau singularities. This restriction is partly by choice and partly by necessity: non-topological
data are much harder to control and compute.
There is one more valuable piece of gauge theory data, however,
that can be extracted with precision from this geometric perspective, namely the holomorphic superpotential $W$.

The superpotential in the quiver gauge theories for the type IIB
D-branes is a holomorphic quantity, and does not depend on the
K\"{a}hler moduli. Hence one can go to the large volume limit and find it from
the topological B-model.
Some superpotentials for the quiver gauge theories on the cones over
dell Pezzo surfaces can be found e.g. in
\cite{Martijn-large-volume}.

For quiver gauge theories, the superpotential is a sum of gauge invariant traces over ordered products of bi-fundamental chiral fields. There is one such term for each oriented closed loop on the  quiver.
Via the superpotential $W$ we can adorn the quiver with additional structure, namely a set of relations
between the bi-fundamental fields given its critical points:
\be
{\partial W \over \partial \phi^a} = 0
\ee
%Here we will gie a short outline of how to compute $W$ from a given CY singularity.
%First, let us discuss its structure for the second example of the previous section.
For a given D3-brane configuration on a  Calabi-Yau singularity, one can compute
$W$ as follows.
For simplicity, let us assume that the closed loops in the quiver are all triangles,  so that $W$ is
a purely cubic function%
\footnote{This is true for the three block exceptional collections
that we discuss below.}
\be
W = C_{abc}{\rm Tr}( \phi^a \phi^b \phi^c)
\ee
%As we have seen, this restriction still leaves us with an interesting class
%of examples.
Now suppose we want to compute the cubic coupling of the chiral multiplets
running between the fractional branes  $\FF_i$, $\FF_j$ and $\FF_k$. Geometrically,
the chiral fields are elements of the Ext groups between the respective sheaves.
If we give ourselves the freedom to choose any basis of chiral fields, we can pick any favorite
set of generators of these Ext-groups and compute their so-called Yoneda pairings by taking the
product of the first two sets of generators
\bea
{\rm Ext} ^l(\FF_i,\FF_j) \times {\rm Ext}^m(\FF_j, \FF_k)  \to {\rm Ext}^{3-n}(\FF_i,\FF_k).
\eea
and decompose the result in terms of the third basis of generators.  Here we used that the
cubic pairing between three Ext groups is non-vanishing only if  $n+l+m=3$, and that
Ext${}^{3-n}(F_i,F_k)$ is the natural dual space to Ext${}^n(\FF_k,\FF_i)$. This calculation
was done for del Pezzo singularities with $n\leq 5$ in \cite{Martijn-large-volume}.

The superpotential resulting from this calculation is a holomorphic function of the space of
chiral bi-fundamental fields, as well as on the space of complex structure deformations
of the del Pezzo singularity. For the $n$-th del Pezzo surface this amounts to
$2(n-4)$ complex parameters, corresponding to the positions of the $n-4$ blow up points
that can not be held fixed by using the $SL(3,{\bf C})$ isometry  group of the underlying
$\PP^2$.
%In the "classical" limit the
%superpotential depends on the complex deformations of the
%geometry.
More generally the superpotential depends also on the
non-commutative deformations and
'gerbe' deformations \cite{Wijnholt-paramet}.

\bigskip

{\small

\noindent
{\bf Example: D3 at the tip of the cone over
$\PP^1\times\PP^1$.}

The classical superpotential depends on the complex deformations of
the cone.
Since there are no deformations of $\PP^1\times\PP^1$ the potential
will have no free parameters.
In order to find the terms in the superpotential,
one, first, takes the fields around a loop in the quiver, this is necessary
for the term to be gauge invariant, then finds the decomposition of the
corresponding homomorphisms in a direct sum and picks up the terms
proportional to identity.
For example, the composition of the homomorphisms
for the upper left triangular in figure 2
reads
\bea
\Hom(\FF_2,\FF_1)\circ\Hom(\FF_4,\FF_2)\circ\Hom(\FF_1,\FF_4)\qquad \quad \\[2.5mm]
=(B_kw^*_k)\circ(A_lz^*_l)\circ(E_{ij}z_iw_j)
\; =\; E_{ij}A_iB_j+\ldots
\eea
where we assume the summation over all repeated indices and the dots
denote the terms not proportional to identity operator.
Taking into account the lower left triangular we get the
superpotential
\be
W=\sum_{i,j=1,2}(A_iB_j-C_iD_j)E_{ij}
\ee
In order to get some intuition one can also compare this calculation
with the orbifold calculation, where the terms in the superpotential
consist of the fields around loops
\cite{Lawrence:1998ja}.

The classical complex deformations of the geometry are not the only
source of deformations of the superpotential. Non
commutative deformation can also play a role.
%In order to motivate this
Let us count the number of possible
parameters in the superpotential  for the quiver in figure 2.
There are $32$ gauge invariant combinations of the fields and the
superpotential may have the form
\be
W=\ld_{ijkl}A_iB_jE_{kl}
+\td\ld_{ijkl}C_iD_jE_{kl}
\ee
%In the large volume limit we neglect the Kahler potential,
We may allow any $Gl(2)$ transformations of the fields $A_i$,
$B_i$, $C_i$, $D_i$ and any $Gl(2)\otimes Gl(2)=Gl(4)$
transformations of $E_{ij}$.
The resulting number of reparameterizations of the fields is
$4\cdot 4+16=32$, but there are three rescalings that don't change the
superpotential: the first one is $A\ra \al_1 A$ and $B\ra \frac{1}{\al_1}B$;
the second one is $C\ra \al_2 C$ and $D\ra \frac{1}{\al_2}D$;
the third one is $E\ra \al_3^2E$ and
$(A,B,C,D)\ra \frac{1}{\al_3} (A,B,C,D)$.
The superpotential thus has three deformation parameters.
And, indeed, there are non commutative deformations of the geometry
given by the inverse $B$-field with holomorphic indices that give
the three-parametric deformations of the superpotential
\cite{Wijnholt-paramet}.
}

\bigskip

\subsection{Kahler potential}

The Lagrangian for the quiver gauge theory
can be deduced from the Born-Infeld and Chern-Simons
parts of the action for the branes
\bea
\label{dbi}
\mathcal{S} =
%\mathcal{S}_{\text {BI}}
%\, + \,
%\mathcal{S}_{\text {CS}}\, , \nonumber
%\\[5mm]
%\mathcal{S}_{\text {BI}}\is
\int\!
e^{-i^*_s \phi}\sqrt{{\rm det}(i_s^*\, G - {\cal F}_s)} \,
%\\[3mm]
%\mathcal{S}_{\text{CS}}\is
+ \int\!
\sump\; i_s^* C_p \; e^{{\cal F}_s}\, ,
\eea
\be
{\cal F}_\sss = F_s -i_s^*B \, . \qquad
\ee
The parameters in this Lagrangian depend on the expectation values
of the background fields corresponding to the closed string modes,
such as the metric on the internal space, the NSNS
B-field, and the RR fields.
Here $i^*_s$ denotes the pull-back of the various fields to the
world-volume of the branes.

%Now that we know how to find the field content of quiver gauge
%theories it is the time to discuss their lagrangians.

%To start let us briefly discuss how many closed string modes
%enter the low energy lagrangian.
%Clearly the modes should be massless, thus the question is how many
%massless modes are there
%for the deformation of the metric, the NSNS B-field and the RR C-fields.

Massless fields arising from type
IIB superstrings compactified on a $CY_3$ fold are organized in $\N=2$
multiplets. In general, there are \cite{Vafa:1997pm}
\\[1mm]
 ${}$  \quad $\bullet$ $h^{1,1}+1$ hypermultiplets\\[1mm]
  ${}$  \quad $\bullet$ $h^{2,1}$ vector multiplets\\[1mm]
  ${}$ \quad  $\bullet$ 1 tensor multiplet\\[1mm]
If we introduce a D-brane, then the $\N=2$ supersymmetry gets broken
to $\N=1$.
The $\N=2$ hypermultiplets split into two sets of chiral multiplets.
One set of these multiplets corresponds to the holomorphic
couplings for the gauge groups
\be
\tau=\frac{\theta}{2\pi}+i\frac{4\pi}{g^2}
\ee
which enter the kinetic terms for the gauge fields on the branes
\bea
{\rm Im}\!\int \! d^2\theta\frac{\tau}{8\pi}W_\al W^\al=
-\frac{1}{4g^2}F\wedge *F+\frac{\theta}{32\pi^2}F\wedge F
\eea
The other field is a holomorphic extension of the FI parameter
\be
S=\rho+i\zeta.
\ee
If the corresponding closed string modes have normalizable kinetic
terms, then $S$ is a dynamical field and
it is possible to write a gauge invariant mass
term for the gauge field on the brane
\cite{Douglas:1996sw}\cite{Dine:1987xk}
\ba\lb{stuck}
\int d^4\theta \frac{1}{4}({\rm Im}(S-\bar S-2V))^2 \qquad
\qquad \qquad \qquad & &\\[-.5mm]
\nonumber
=\frac{1}{2}(A-d\rho)\wedge *(A-d\rho)-\zeta D & &
\ea
consequently $\rho$ is interpreted as the Stuckelberg field
and $\zeta$ is indeed the FI  `parameter'.
In the non compact geometry the normalizable modes correspond to
Poincare dual cycles that are both compact.
For the cone over del Pezzo surfaces, such cycles are the four
cycle and the canonical class.
The branes wrapping these compact Poincare dual cycles have chiral
matter in their intersections, i.e. the corresponding $U(1)$ gauge
theories on their world volume have mixed anomalies.
These anomalous $U(1)$ gauge groups become massive due to the
interaction with the normalizable modes of the closed strings via
interaction (\ref{stuck}).

If the string modes corresponding to $S$ are non normalizable, then
the fields $\rho$ and $\zeta$ become parameters and the only gauge
invariant combination is $\zeta D$, i.e. the usual FI parameter of
the gauge theory.
Although the $U(1)$ fields interacting with $S$ don't have anomalies
in this case, they can still get a mass through the Higgs mechanism
induced by non zero FI parameter $\zeta$.

Expanding the DBI and the Chern-Simons actions one finds the
following gauge couplings for the D3 and D5 branes%
\footnote{The expressions for the D7 brane can be derived in a
similar way.
}
\bea
\tau_0=C_0+ie^{-\vp}\qquad \ \\[1mm]
%while the gauge theory coupling for the D5-brane
\tau_2=\int_{\PP^1}(C_2+\tau_0 B_2)
\eea
Stuckelberg field and FI parameter for the D5-brane are
\ba
\nonumber
d\rho=*_4d\int_{\PP^1}C_4\, , \qquad \qquad
\zeta=\int_{\PP^1} J\, .
\ea
In general,
the stable branes are represented by the bound states that have D3,
D5 and D7 charges. Hence the coupling on these fractional branes
will be a combination of the above couplings. It can be shown
that the gauge coupling associated to a fractional BPS brane with the charge
vector $Q$ can be obtained from the central charge vector via
\be
\frac{4\pi}{g^2}=e^{-\vp}|Z(Q)|
\ee
while the FI parameter for the gauge theory on the fractional brane is
proportional to the phase of the central charge
\cite{Douglas:2000gi}
\be
\zeta={1\over \pi} {\rm Im}\log Z(Q)
\ee
The intuitive motivation for this identification is that two
D-branes with different phases of their central charges break the
supersymmetry in a similar way as FI parameters do.
A formal derivation of this correspondence
in terms of the SCFT on the branes can be
found e.g. in \cite{Douglas:2000gi}.

\bigskip

{\small
\noindent
{\bf Example: D3 near a $Z_2$ orbifold point.}\\
In this example we find the gauge couplings for the fractional
branes on the $Z_2$ orbifold singularity.
The fractional branes near a resolved $Z_2$ orbifold have the central charges
\be\lb{cencha}
Z=Q_3+Q_5\int_{\PP^1} (B+iJ)
\ee
In the orbifold limit $J=0$.
The charges $(Q_5,\:Q_3)$ of the fractional branes are $(-1,\:1)$
and $(1,\:0)$.
The corresponding couplings are
\be\lb{coupl}
\frac{4\pi}{g_1^2}=e^{-\vp}\left(1-\int B\right)\;\;\;\;\;\;\;
\frac{4\pi}{g_2^2}=e^{-\vp}\int B
\ee
For the orbifold $\int B=1/2$ \cite{Aspin-orbi} and the couplings are equal.
It is interesting to note that the more general expressions
(\ref{coupl}) work also for the conifold \cite{Klebanov:2000hb}.
This is not too surprising because the conifold can be obtained from
the $Z_2$ orbifold by giving certain masses to the adjoint
fields \cite{Klebanov:1998hh}.

The blow up of the two-cycle at the $Z_2$ singularity corresponds to
$J\neq 0$.
In this case the absolute value of the central charges
(\ref{cencha}) is bigger than $1/2$, consequently it becomes more
favorable for the fractional branes to recombine in D3-branes.
From the quiver gauge theory point of view,
the resolution of the singularity can be reproduced by the
non zero FI parameters
\cite{Douglas:1996sw}\cite{Kronheimer:1989zs}.
In the presence of FI parameters,
some of the bifundamental fields get VEVs and
break the gauge group $U(N)\times U(N)\lra U(N)_{diag}$.
This corresponds to the recombination of two stacks of fractional
branes in one stack of D3-branes.
The massless fields that solve the F and D-term equations represent the
motion of these D3-branes on the resolved manifold.
}

\bigskip

The coupling of the gauge fields to the RR-fields follows from expanding the CS-term of the action.
In this way we derive that the $\theta$-angle has the geometric expression
\ba
%\nonumber \\[4mm]
\theta =Q_7\int_{\PP^2} C^{(4)}+%\sum_A
Q^A_5\int_{\al_A} \!\! C^{(2)}
+Q_3 \:C^{(0)}
% Q_4 \theta_{4} + Q_2^A \theta^A\, + Q_0 C^{( 0)}   , %( p_{s\aaa} - r_s b_\aaa)%
%\qquad \quad \theta^\aaa  = \! \int_{\alpha_\aaa} \! C_{\it 2}\, .
\ea
%with $\theta^\aaa$ as defined in (\ref{taua})
%The gauge theory on $\FF_s$
In addition, each fractional brane may support a St\"uckelberg field, which arises by dualizing  the RR
2-form potential ${\mathcal C}$ that couples linearly to the gauge field strength via
\be
\label{linear}
{\mathcal C} \wedge F
\ee
From the CS-term we read off that
\bea
\label{pot}
{\mathcal C}
=Q_4\int_{\PP^2} C^{(6)}+%\sum_A
Q^A_2\int_{\al_A} \!\! C^{(4)}
+Q_0 \:C^{(2)}
\eea
The linear coupling (\ref{linear}) determines the spectrum of $U(1)$ gauge bosons, as we will now show.
\bigskip

\noindent
\subsection{Spectrum of $U(1)$ Gauge Bosons}

On $dP_n$ there are $n+3$ different fractional branes, with a priori as many independent gauge
couplings and FI parameters. However, the  above expression for the central charge contain
only $2n+4$ independent continuous parameters: the dilaton, the (dualized) B-field,
and a pair of periods (of $B$ and of $J$) for every of the $n+1$ 2-cycles in $dP_n$.
Hence there must be two relations restricting the couplings. The interpretation of these
relations is that quiver gauge theory always contains two anomalous
$U(1)$ factors. FI-parameters associated with anomalous $U(1)$'s are not freely tunable,
but dynamically adjusted so that the associated D-term equations are automatically satisfied.
This adjustment relates the anomalous FI variables and gauge couplings.

The non-compact cone $Y_0$ supports two compact cycles for which the dual cycle is also
compact, namely,  the canonical class and the del Pezzo surface $X$. Correspondingly, we
expect to find normalizable 2-form and 4-form on $Y_0$.
Their presence implies that two closed string modes survive as dynamical 4-d fields with normalizable
kinetic terms; these are the two axions associated
with the two anomalous $U(1)$ factors. The two $U(1)$'s are dual to each other:
a $U(1)$ gauge rotation of one generates an additive shift in the $\theta$-angle of the other.
This naturally identifies the respective $\theta$-angles and St\"uckelberg fields.
The geometric origin of this identification is that the corresponding
branes wrap dual intersecting cycles.

We obtain non-normalizable harmonic forms on the non-compact cone
by extending the other harmonic 2-forms  on $X$ to $r$-independent forms.
The corresponding 4-d RR-modes are non-dynamical fields:
any space-time variation would carry infinite kinetic
energy. Thus for the non-compact cone $Y$, all non-anomalous $U(1)$
factors remain massless.

As we will show below, the story changes for the compactified setting,
for D-branes at a del Pezzo singularity inside a compact CY threefold $Y$.
In this case, a subclass of all harmonic forms on the cone may extend to
normalizable harmonic forms on $Y$, and all corresponding closed string modes
are dynamical 4-d fields. This may lead to mass terms for the non-anomalous $U(1)$'s.

Consider IIB string theory compactified on
a Calabi-Yau orientifold $Y$. The relevant RR fields
%$\mathcal{O}=(-1)^{F_L}\Omega_p\sigma$,
decompose into appropriate harmonic forms on $Y$ as follows
\cite{Grimm:2004uq}
\ba
\label{c6}
C^{(6)}\is {\ccc}_{a}(x)\, {\omega}^a \,
\nonumber \\[2.5mm]
C^{(4)}\is % \raisebox{.4mm}{$\sum\limits_\alpha$} \, (
 \ccc^{\, \alpha}(x)\, \omega_{\alpha}
+\, \rho_{\alpha}(x)\, {\omega}^{\alpha} \\[2.5mm]
C^{(2)}\is  % \!\! \raisebox{.4mm}{$\sum\limits_a$} \;
 \theta^{a}(x)\,\omega_a\, . \nonumber
\ea
Here $\omega^a$ and $\omega^\alpha$ are harmonic 4-forms on $Y$, and respectively
even and odd under the orientifold projection. Similarly, $\omega_a$ and $\omega_\alpha$
are (even and odd) harmonic 2-forms.
So in space-time, $\ccc^{\alpha}$, $\ccc_a$, and $\ccc_{\it 2}$ are two-form fields
and $\rho_\alpha$ and $\theta^a$ are scalar fields.
Similarly, we can expand the K\"ahler form $J$ and NS B-field as
\ba \label{b2}
J= \zeta^{\alpha}(x) \, \omega_{\alpha}\, \\[2mm]
\nonumber
B = b^{a}(x) \, {\omega}_a
\ea
Note that the orientifold projection,
in particular,
eliminates the constant zero-mode components of $C^{(2)}$, $C^{(6)}$ and $B^{(2)}$.
The IIB supergravity action
contains the following kinetic terms for the RR
$p$-form fields
\ba
\label{bulkaction}
\mathcal{S} \is
\int [ \, G^{ab}\, d\ccc_a\! \wedge\! *
d\ccc_b \, + \, %{\mathcal{K}^2}
G_{\alpha\beta}\, d\ccc^{\alpha} \! \wedge\! * d\ccc^{\beta}\, ]
\ea
where $G_{\alpha\beta}$ and $G^{ab}$
denote the natural metrics on the space of harmonic 2-forms on $Y$
\bea\label{condforms}
%\int_Y\omega_{\alpha}\wedge{\omega}^{\beta} = \delta^{\beta}_{\alpha},
%\qquad & &
%\int_Y\tilde\omega_{a}\wedge\tilde{\omega}^{b} = \delta^{b}_{a}\, \, \\[2mm]
G_{\alpha\beta}= %\raisebox{-.5mm}{\Large $\frac{1}{\mathcal{K}}$}
\int_Y\omega_{\alpha}\wedge*
\omega_{\beta}
\qquad & &
G^{ab} = %\raisebox{-.5mm}{\Large $\frac{1}{\mathcal{K}}$}
\int_Y\omega^a\wedge* \omega^b %\\[3mm]
%{\mathcal K} \is \int_Y  J\wedge J\wedge J \nonumber
\eea
%and ${\cal K}$  denotes the K\"ahler volume of $Y$
The scalar RR fields $\theta^b$ and $\rho_\alpha$ are related to the above 2-form
fields via the duality relations:
\bea\label{cctildedual}
*\, d\theta^{\, b} = -G^{ab} d \ccc_a\, , \qquad
*\, d\rho_{\alpha} = %{\mathcal{K}^2}
G_{\alpha\beta} d\ccc^{\beta}\, .
\eea
The harmonic forms on the compact CY manifold $Y$, when restricted to base $X$ of the
singularity, in general do not span the full cohomology of  $X$.  For instance, the
$(1,1)$-cohomology of $Y$ may have fewer generators than that
of $X$, in which case there must be one or more 2-cycles that are non-trivial within $X$
but trivial within $Y$. Conversely, $Y$ may have non-trivial cohomology elements
that restrict to trivial elements on $X$. The overlap matrices
\ba
\label{pi}
\Pi_{\alpha}^\aaa \is \int_{\alpha_\aaa} \omega_\alpha \, ,%\nonumber
\qquad \qquad %\\[3mm]
\Pi^{\aaa}_a \, = \, \int_{\alpha_\aaa} \omega_a  \, ,
\ea
when viewed as linear maps between cohomology spaces $H^{(1,1)}(X,\Z)$ and
$H^{(1,1)}_{\pm}(Y,\Z)$, thus typically have both non-zero kernel and cokernel.

This incomplete overlap between the two cohomologies has immediate repercussions
for the D-brane gauge theory, since it implies that the compact embedding typically reduces
the space of gauge invariant couplings. The couplings are all period integrals of certain
harmonic forms, and any reduction of the associated cohomology spaces reduces the
number of allowed deformations of the gauge theory. This truncation is independent from
the issue of moduli stabilization, which is a {\it dynamical} mechanism for fixing
the couplings, whereas the mismatch of cohomologies amounts to a {\it topological}
obstruction.

By using the overlap matrices (\ref{pi}), we can expand the  topologically available
local couplings in terms of the global periods, defined in (\ref{c6}) and (\ref{b2}), as
\ba
b^{\, \aaa} = \Pi^\aaa_{\; a}\, b^a\, , \quad\ &  &\ \quad
\ccc^{\, \aaa} =  \Pi^{\aaa}_{\; \alpha}\,
\ccc^{\, \alpha}\nonumber \\[3mm]
 \theta^{\, \aaa} =  \Pi^{\aaa}_{\; a}\, \theta^{\, a} \, , \quad\ & &\ \quad
\zeta^{\, \aaa} =  \Pi^{\aaa}_{\; \alpha}\,
\zeta^{\alpha} \nonumber
\ea
By construction, the fields on the left hand-side are
elements of the subspace of $H^{(1,1)}$ that is
common to both $Y$ and $X$. The number of independent closed
string couplings of each type thus coincides with the rank of the corresponding overlap matrix.

\newcommand{\two}{{\mbox{\scriptsize 2}}}
\newcommand{\sums}{\mbox{\large $\sum\limits_{\raisebox{.5mm}{\scriptsize $s$}}$}}

As a special consequence, it may be possible to form
linear combinations of fractional branes, such that the charge
adds up to that of a D5-brane wrapping a 2-cycle
within $X$ that is trivial within the total space $Y$. The D5-brane charge for such a
linear combination of branes satisfies
\be
\Pi^\aaa_\alpha Q_5^\aaa = 0
\ee
for all $\alpha$. As a result, the corresponding
$U(1)$ vector  boson decouples from the normalizable RR-modes,
and thus remains massless. This observation will have an important application in the
next section.

Let us compute the non-zero masses.  Upon dualizing, or equivalently, integrating out the 2-form potentials, we obtain (among other terms)
a St\"uckelberg mass term for the vector bosons $A_s$ of the form
\ba
\label{finalmass}
%G_{{}_{\! XX} } \,\nabla \rho^{{}_X} \wedge * \nabla
%\rho^{{}_X}
%+ \,
G^{\alpha\beta} \,
\nabla \rho_\alpha \wedge * \nabla \rho_\beta
\ea
%\vspace{-4mm}
with
\ba
%\nabla \rho^{{}_X} \is d \rho^{{}_X}
%- \sums
%\, \rr_{\sss} \, A_{s}\, , \nonumber \\[3mm]
\nabla \rho_\alpha \is d\rho_\alpha -
%\sums
 \; Q_5^\aaa
\, \Pi^\aaa_{\, \alpha}
\, A \, .
\ea
The vector boson mass matrix
%\be\tau^0 = \rho_0 + i \int _X\! J \wedge J  \ee
\be
m^2 = % G_{{}_{\! XX}}\, \rr_s \rr_{s'} +\,
 G^{\alpha\beta}
\,\Pi^{\; \aaa}_{\alpha}
\Pi^{\; \bbb}_{\beta} \,
Q_5^{\aaa} Q_5^{\bbb}
\ee
is of order of the string scale for string size compactifications.\footnote{The
$U(1)$ masses can be made much lower than the string scale by considering compactifications
with large extra dimensions. For our discussion, however, we assume that the compactification
manifold is of string size.}
It lifts all $U(1)$ vector bosons from the low energy spectrum, except for the ones
that correspond to fractional branes that wrap 2-cycles that are trivial within $Y$.
This is the central result of this subsection.

\bigskip

\noindent
\subsection{Symmetry breaking}

\smallskip

The quiver gauge theory on a D3-brane at a CY singularity
contains a number of $U(1)$ factors, one for
each type of fractional branes.
For each non anomalous $U(1)$, one can turn on an FI-parameter $\zeta_i$.
The  FI-parameters typically correspond to blow-up modes that govern the size
of two-cycles within the CY manifold.

A more precise correspondence
can be extracted by studying the D and F flatness equations that select the supersymmetric
classical vacua of the ${\cal N}=1$  gauge theory. This space of vacua can be thought
of as the configuration space of the D3 brane within the CY singularity. The F-flatness
conditions follow from extremizing the holomorphic superpotential,  $\partial W/\partial \phi^a = 0$,
for all chiral matter fields $\phi^a$. The D-term equations further restrict this space of solutions.
There is one real D-flatness condition for each node $i$ on the quiver
\be
\label{dflat}
\mu_i \equiv \sum_{a}  |\phi^+_{ia}|^2 - \sum_b |\phi^-_{ib}|^2 - \zeta_i = 0,
\ee
where $\phi^+_a$ and $\phi^-_b$ denote the bi-fundamentals on each side of the
corresponding node.  The conditions (\ref{dflat}) are implemented via
a symplectic quotient: it identifies field configurations on a gauge orbit generated by
$\delta \phi^\pm_{ia}  = {\pm \epsilon}  \phi_{ia}$ and sets $\mu_i = 0$.
Each D-constraint thus eliminates two dimensions from the solution
space of the F-flatness equations.

For a given quiver gauge theory associated to a  D3-brane on a Calabi-Yau singularity,
the moduli space of vacua, the space of solutions to the F- and D-term equations,
reconstructs the geometry of the CY singularity. This correspondence may provide an
interesting route towards reverse engineering the appropriate CY geometry associated
to a given quiver gauge theory. In general, however, quiver gauge theories do not lead to
simple commutative geometries.

By varying the ambient Calabi-Yau geometry, fractional branes can become unstable:
they may decay into two or more components  or
form bound states.  In the large volume theory, this happens because the
central charges of the fractional branes may re-orient themselves such that
the mutual triangle inequalities, that ensure their stability, get violated.
From the perspective of the quiver gauge theory, the formation of
a bound state is described by condensation of one or more bi-fundamental scalar
fields. This generically happens as soon as some of the FI-parameters are non-zero.
The D-term equations (\ref{dflat}) then dictate that some scalar fields $\phi^c$
must acquire a non-vanishing vacuum expectation value $\langle \phi^c\rangle$.
In the new vacuum, part of the gauge symmetry gets broken. In addition, several
matter fields acquire a mass proportional to $\langle \phi^{c}\rangle$  and get
lifted from  the moduli space of supersymmetric vacua via the F-flatness equation
\be
\label{lift}
{\partial W \over \partial \phi^a} = C_{abc} \, \phi^b  \, \langle \phi^c\rangle  = 0
\ee
Hence the number of fields that become massive is determined by the number
of non-zero Yukawa couplings $C_{abc}$ of the field $\phi^c$ that acquires the vev.

\bigskip

\noindent
\section{Bottom-Up String Phenomenology}

The basic strategy for finding D-brane realizations of realistic gauge theories proceeds via a two step process. First one looks for CY singularities
and brane configurations, such that the quiver gauge theory is just rich enough to contain the SM gauge
group and matter content.
Then we look for a well-chosen symmetry breaking process that
reduces the gauge group and matter content to that of the Standard Model, or at least
realistically close to it.
When the CY singularity is not isolated, the moduli space of vacua for
the D-brane theory has several components \cite{Morrison:1998cs},
and the symmetry breaking we need is found on a component in which some of
the fractional branes move off of the primary singular point along
a curve of singularities (and other branes are replaced by appropriate
bound states).  This geometric insight into the symmetry
breaking allows us to identify an appropriate CY singularity, such that
the corresponding D-brane theory
looks like the SSM.

The above procedure was used in \cite{Verlinde:2005jr} to construct a semi-realistic theory
from a single D3-brane on a partially resolved del Pezzo 8 singularity. The final model,
however, still had several extra $U(1)$ factors besides the hypercharge symmetry.
Such extra $U(1)$'s are characteristic of D-brane constructions: typically, one
obtains one such factor for every fractional brane. Which of these $U(1)$ factors remains
massless depends on the embedding  of the CY singularity inside a full compact CY geometry.
As we have seen, the massless $U(1)$ gauge bosons are in one-to-one
correspondence with non-trivial 2-cycles
within the local CY singularity that lift to {\it trivial} cycles within the full CY three-fold.
This insight can be used to ensure that, among all
$U(1)$ factors, only the hypercharge survives
as a massless gauge symmetry.

The interrelation between the 2-cohomology of the del Pezzo base of the singularity, and
the full CY thee-fold has other relevant consequences.
Locally, all gauge invariant couplings of the D-brane theory can be varied  via
corresponding deformations of the local geometry. This local tunability is one of the
central motivations for the bottom-up approach to string phenomenology. The embedding into a
full string compactification, however, typically introduces a topological obstruction against
varying all local couplings: only those couplings that descend from moduli of the full CY survive.
Their value will need to be fixed via a dynamical moduli stabilisation mechanism.

% \bigskip \noindent
 %\subsection{Summary}

Let us summarize our general strategy: % We formulate it as a systematized set of steps:

\begin{itemize}
\item{Choose a non-compact CY singularity, $Y_0$,
%Make a list of all Calabi-Yau singularities $Y_{0}$
and find a suitable basis of fractional branes $\FF_i$ on $Y_{0}$.
Assign multiplicities $n_i$ to each  $\FF_i$ and enumerate the resulting quiver gauge theories.}
\item{Look for quiver theories that, after symmetry breaking, produce an SM-like theory.
Use the geometric dictionary to identify the corresponding resolved CY singularity.}
\item{Identify the topological condition that isolates hypercharge as the only massless $U(1)$.
Look for a compact CY 3-fold $Y$, with the right topological properties, that contains ${Y}_0$.
Use fluxes and other ingredients to stabilize the moduli of $Y$ at the desired values.}
\end{itemize}
\smallskip

\noindent
In principle, it should be possible to automatize several of these steps and thus set up
a computer-aided search of Standard Model constructions based on D-branes at CY singularities.
%For this paper, we have not done this.

\bigskip

\noindent
\subsection{A Standard Model D-brane}

We now apply the lessons of the previous section to the string construction of a Standard
Model-like theory of \cite{Verlinde:2005jr}, using the world volume theory of a D3-brane on a del Pezzo 8
singularity. Let us summarize the set up. % -- more details are found in \cite{MH}.

Mathematicians have identified a conventient class of exceptional collections,
or bases of fractional branes, on a del Pezzo 8 singularity. These bases have several
desirable characteristics. In particular, all fractional branes in an exceptional collection
wrap so-called rigid cycles, with cohomological properties that eliminate translational modes.
In terms of the world-volume gauge theory, this ensures the absence
of adjoint matter besides the gauge multiplet. All charged matter appears in the form of
bi-fundamentals, that live on the brane intersections.

 The construction of \cite{Verlinde:2005jr} starts from a single
D3-brane; the multiplicities $n_i$ are then uniquely determined via the condition (\ref{decos}).
For a favorable basis of fractional branes,
this leads to an ${\cal N}=1$ quiver gauge theory with the gauge group
$
{\cal G}_0 = U(6) \times U(3) \times U(1)^9.
$
This particular quiver theory is  related via a single Seiberg duality to the world volume theory of
a D3-brane near a  $\C^3/\Delta_{27}$ orbifold singularity -- the model considered earlier
in \cite{delta27} as a possible starting point for a string realization of a Standard Model-like
gauge theory.  As  shown in \cite{Verlinde:2005jr}, the theory with the above gauge
group ${\cal G}_0$ is rich
enough, so that one can design a symmetry breaking process to a semi-realistic
gauge theory with the gauge group
$$
{\cal G} =U(3) \times U(2) \times U(1)^7.
$$
\noindent
The quiver diagram is drawn in fig 3.

\begin{figure}[t]
\begin{center}
%\leavevmode\hbox{\epsfxsize=7cm \epsffile{finquiver.eps}}\\[3mm]
\epsfig{figure=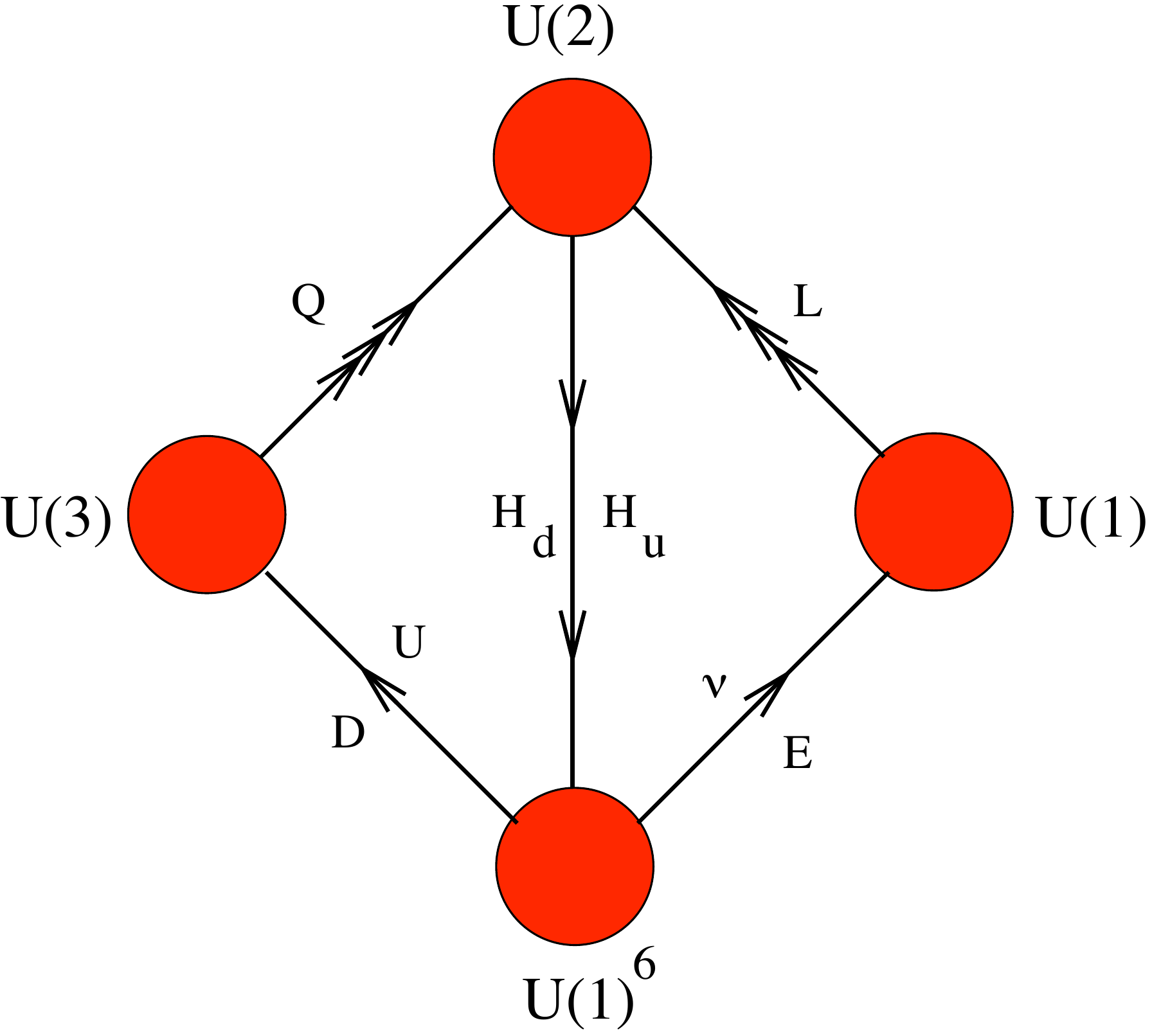,scale=0.38}
\vspace{-8mm}
\caption{The MSSM-like quiver gauge theory obtained in \cite{Verlinde:2005jr}.
Each line represents three generations of bi-fundamentals.
% In the text below we will identify the geometric condition that isolates
%the $U(1)_Y$ hypercharge as the only surviving massless $U(1)$ gauge symmetry.
}
\vspace{-8mm}
\label{East1b}
\end{center}
\end{figure}

Each line represents three generations of
bi-fundamental fields.  The D-brane model thus has the same
non-abelian gauge symmetries, and the same quark and lepton content as the Standard Model.
It has an excess of Higgs fields -- two pairs per generation -- and several
extra $U(1)$-factors. Our plan is to use the new insights obtained in section 5.3
to move the model one step closer to reality, by eliminating all the extra $U(1)$ gauge
symmetries except hypercharge from the low energy theory.

The strategy for producing the gauge theory of fig \ref{East1b}
is this: by appropriately tuning the superpotential
(i.e., varying the complex structure) we can find a Calabi--Yau with
a non-isolated singularity---a curve $\Gamma$ of $A_2$ singular points---such
that the classes $\alpha_1$ and
$\alpha_2$ have been blown down to an $A_2$ singularity on the (generalized)
del Pezzo surface where it meets the singular locus. The symmetry-breaking
involves moving onto the $\Gamma$ branch in the moduli space, where
the $\alpha_1$ and $\alpha_2$ fractional brane classes are free to
move along the curve $\Gamma$ of $A_2$ singularities.  In particular,
these branes can be taken to be very far from the primary singular point
of interest, and become part of the bulk theory: any effect which they
have on the physics will occur at very high energy like the rest of
the bulk theory.

Making this choice removes the branes supported on $\alpha_1$ and
$\alpha_2$ from the original brane spectrum, and replaces other
branes in the spectrum by bound states which are independent of $\alpha_1$ and
$\alpha_2$.
The remaining bound state basis of the fractional branes
is specified by the following set of charge vectors
\bea
\label{coll2}
{\rm ch}(\FF_{1})  = (3, -2K + \! \mbox{$\sum\limits_{i=5}^8 E_i$}\! - E_4,  {\textstyle{1\over 2}} )
\qquad \qquad \qquad \ \ \\[-.5mm]
\, {\rm ch}(\FF_{2}) =\, (\, 3,\, \mbox{$\sum\limits_{i=5}^8 E_i$},  \textstyle{-2})\qquad \qquad \qquad \qquad \qquad \  \ \; \\[-.5mm]
{\rm ch}(\FF_{3})  = \,  (3,3H-\! \mbox{$\sum\limits_{i=1}^4 E_i$}, -\textstyle{ 1\over 2} )\qquad \qquad \qquad \qquad  \ \ \\[2mm]
{\rm ch}({\FF}_{4})\, = \, (1, H -E_4, 0) \ \  \; \qquad \qquad \qquad\qquad \qquad \\[2.9mm] {\rm ch}(\FF_{i})  \, = \, (1,- K\! + E_i , \,1\, ) \qquad \mbox{\scriptsize ${i =5,\, .  \,   ,8}$}\qquad \ \ \,  \qquad \\[2mm]
{\rm ch}({\FF}_{9}) \, =  \, (1, 2H - {\mbox{$\sum\limits_{i=1}^4$}} E_i, 0) \qquad\qquad \quad  \qquad \qquad \;
\eea
Here the first and third entry indicate the D7 and D3 charges; the second
entry gives the 2-cycle wrapped by the D5-brane component of $\FF_i$.
As shown in \cite{Verlinde:2005jr}, the above collection of fractional branes is rigid, in the sense that
the branes have the minimum number of self-intersections and the corresponding gauge
theory is free of adjoint matter besides the gauge multiplet.
{}From the collection of charge vectors, one easily obtains the matrix of intersection
products via the fomula (\ref{int}). One finds
\bea
\label{chiminus}
\mbox{\footnotesize $\#(\FF_i,\FF_j)$} =
\mbox{\scriptsize
$\left(\! \begin{array}{ccccccccc}
\ 0 & -3 & \ 0 & 1 &  1 & 1 &  1 & 1 & 1 \\
\ 3 &\ 0 &\ 3 &  2 &  2 & 2 &  2 &  2 &  2 \\
\ 0 & -3 & \ 0 &  1 &  1 & 1 &  1 &  1 & 1 \\
-1 & -2 & -1 & 0 &  0 & 0 &  0 &  0 & 0 \\
-1 & -2 & - 1 &  0 & 0 & 0 &  0 &  0 & 0 \\
-1 & -2 & -1 & 0 &  0 & 0 &  0 &  0 & 0 \\
-1 & -2 & -1 &  0 &  0 & 0 &  0 &  0 & 0 \\
-1 & -2 & -1 &  0 &  0 & 0 &  0 &  0 & 0 \\
-1 & -2 & -1 &  0 &  0 & 0 &  0 &  0 & 0
\end{array}\!
\right) $}
\eea
which gives the quiver diagram drawn in fig 3.  The rank of each gauge group corresponds to the
(absolute value of the)  multiplicity of the corresponding fractional brane, and has been
chosen such that weighted sum of charge vectors adds up to the charge of a single D3-brane.
In other words, the gauge theory of fig 3 arises from a
single D3-brane placed at the del Pezzo 8 singularity.

Note that, as expected, all fractional branes in the basis (\ref{coll2}) have vanishing D5 wrapping
numbers around the two 2-cycles corresponding to the first two roots $\alpha_1$ and
$\alpha_2$ of $E_8$, since we have converted the FI parameters which
were blowup modes for those cycles into positions for $A_2$-fractional
branes. After eliminating the two 2-cycles $\alpha_1$ and $\alpha_2$, the remaining
2-cohomology of the del Pezzo singularity is
spanned by the roots $\alpha_i$ with $i=3, .. , 8$ and the canoncial class $K$.
%ed the charges
%and intersection numbers of the collection of fractional branes $\FF_i$ with the remaining
%2-cycles.
%Note, that after the blowup the branes (\ref{coll2})
%form a complete basis of fractional branes.

\begin{figure}[t]
\begin{center}
%\leavevmode\hbox{\epsfxsize=7cm \epsffile{finquiver.eps}}\\[3mm]
\epsfig{figure=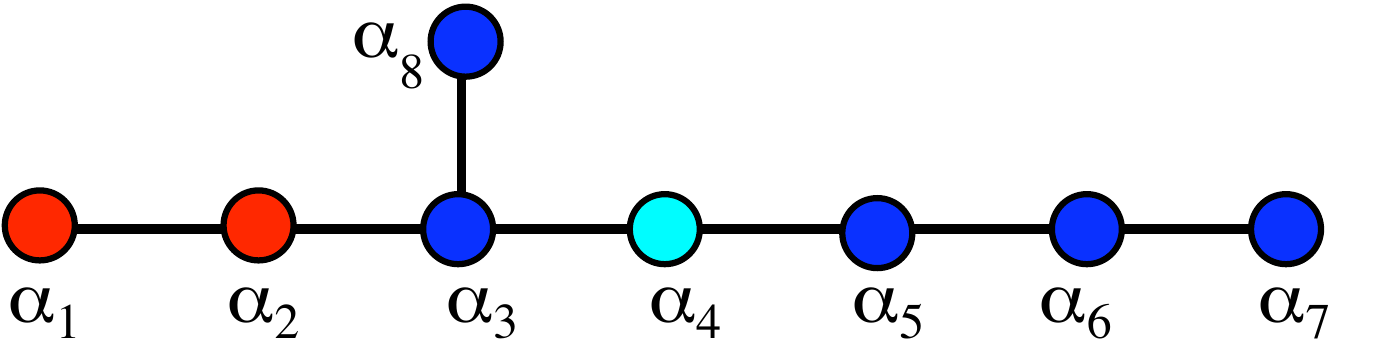,scale=0.5}
\caption{Our proposed D3-brane realization of the MSSM involves a $dP_8$ singularity
embedded inside a CY manifold, such that two of its 2-cycles, $\alpha_1$ and $\alpha_2$, develop an $A_2$ singularity,
and all 2-cycles except $\alpha_4$ are non-trivial within the full CY.}
%\label{default}
\end{center}
\vspace{-5mm}
\end{figure}
\newcommand{\vQ}{\cal Q}
\newcommand{\vL}{\cal L}
\newcommand{\vbQ}{\overline{\vQ}}
\newcommand{\bu}{\overline u}
\newcommand{\bd}{\raisebox{-3pt}{$\overline d$}}
\newcommand{\bbe}{\overline e}
\newcommand{\bq}{\overline q}
\newcommand{\bH}{\overline H}
\newcommand{\bnu}{\overline \nu}

%Ideally, one would like to characterize the geometric deformation of the $$dP_8$$ singularity that
%induces the above symmetry breaking pattern.

%Via the intersection pairing, the 2-cycles form an integral lattice in ${\bf R}^9$.

\subsection{Identification of hypercharge}

Let us turn to discuss the $U(1)$ factors in the quiver of fig 3, and identify the linear combination that defines hypercharge.
We denote the node on the right by $U(1)_1$, and the overall $U(1)$-factors of the $U(2)$ and
$U(3)$ nodes by $U(1)_2$ and $U(1)_3$, resp.  The $U(1)^6$ node at the bottom divides into
two nodes $U(1)^3_u$ and $U(1)^3_d$, where each $U(1)_u$ and $U(1)_d$
acts on the matter fields of
the corresponding generation only.  We denote the nine  $U(1)$ generators
by $\{Q_1, Q_2, Q_3, Q^i_u,Q^i_d, \}$. The total charge
$$
Q_{tot} = \sum_s Q_s
$$
decouples: none of the bi-fundamental fields is charged under $Q_{tot}$.
Of the remaining eight generators, two have mixed $U(1)$ anomalies. As discussed,
 these are associated to fractional branes that intersect compact cycles
within the del Pezzo singularity. In other words, any linear combination of charges
such that the corresponding fractional brane has zero rank and zero degree is free
of anomalies.

\medskip
As seen from table 1, hypercharge is identified with the non-anomalous combination
\be
\label{hyper}
Q_Y =  {1\over 2}  Q_1 - {1\over 6} Q_3 - {1\over 2}
\Bigl( \,  \mbox{$\sum\limits_{{}_{i=1}}^{{}^{{}_3}}$}\,  Q_d^i -  \mbox{$\sum\limits_{{}_{i=1}}^{{}^{{}_3}}$} \, Q^i_u\Bigr)
 \ee
The other non-anomalous $U(1)$ charges are
\ba
{1\over 3} Q_3-{1\over 2} Q_1 \is B-L,
\ea
together with four independent abelian flavor symmetries of the form
\be
Q^{ij}_{u,d} = Q_{u}^i - Q_{u}^j,
\qquad \quad
Q^{ij}_b  = Q^i_b - Q^j_b .
 \ee
 We would like to ensure that, among all these charges, only the hypercharge survives as
 a low energy gauge symmetry.  From our study of the stringy St\"uckelberg mechanism, we
 now know that this can be achieved if we find a
CY embedding of the $dP_8$ geometry such that only the particular 2-cycle associated with $Q_Y$
represents a trivial homology class within the full CY three-fold.
We will compute this 2-cycle momentarily.

{\footnotesize \ba \nonumber
\begin{array}{|c||r|r|r|r|r|r| |c| }
\hline
& \ Q_0\  & \ Q_d \  & \ Q_u\  & \ Q_2\ & \ Q_3\  & \ \ Q_Y \\[1.2mm] \hline \hline
Q & 0 & 0 & 0 & -1 & 1  & \ {1/6} \\[1.2mm] \hline
\bu & 0 & 0 & 1 & 0 & -1 & - {2/ 3} \\[1.2mm] \hline
\bd & 0 & 1& 0 & 0 & -1 & {1/ 3} \\[1.2mm] \hline
L & 1 & 0 & 0 & -1 & 0 & - {1/2}  \\[1.2mm] \hline
\bnu & - 1 & 0 & 1 & 0 & 0 & \ 0 \;  \\[1.2mm] \hline
\bbe & - 1 & 1 & 0 & 0 & 0 & \  1 \; \\[1.2mm] \hline
H^{u} & 0 & 0 & -1 & 1 & 0 &\ {1/2}  \\[1.2mm] \hline
H^{d}  & 0 & -1 & 0 & 1 & 0 & - {1/2} \\[1.2mm] \hline
%g & 0 & -1 & 0 & 0 & 0\\ \hline
%q & 2 & 0 & 0 & 0 & 0 \\
%\hline\bq & -2 & 1 & 0 & 0 & 0\\ \hline
\end{array}\label{eq:charges}
\ea}
\begin{center}
\parbox{9cm}{\small Table 1. $U(1)$ charges of the various matter fields.}
\end{center}

The linear sum (\ref{hyper}) of $U(1)$ charges that defines $Q_Y$,
selects a corresponding linear sum of fractional branes, which we may choose as follows\footnote{\small
With this equation we do not suggest any bound state formation of fractional
branes. Instead, we simply use it as an intermediate step in determining the cohomology
class of the linear combination of branes, whose $U(1)$ generators add up to $U(1)_Y$.}
\be
\FF_{\rm Y} = {1\over 2} \Bigl( \, \FF_{3} -  \FF_0 -\!\! \mbox{$\sum\limits_{{i=4,5,9}}$} \FF_i\, + \! \! \mbox{$\sum\limits_{{i=6,7,8}}$} \FF_i\, \Bigr)
\ee
A simple calculation gives that, at the level of the charge vectors
\be
{\rm ch}(\FF_{\rm Y} ) = (\,0\, , \, -\alpha_4,\, \textstyle{1\over 2} \, ) \qquad \quad \alpha_4 = E_5-E_4
\ee
%Since the D3-charge can be cancelled by a simple shift $Q_Y \to Q_Y - {1\over 2} Q_{tot}$,
We read off that the 2-cycle associated with the hypercharge generator $Q_Y$ is the one represented by the simple root $\alpha_4$.

We consider this an encouragingly simple result.
% It
%means that to lift all other $U(1)$'s, all
%we need is a  compact CY embedding of the $dP_8$ singularity, such that all 2-cycles
 %except $\alpha_4$ lift to non-trivial homology classes. %We now turn to this  challenge.
%\bigskip\bigskip\bigskip\bigskip
 Namely,
when added to the insights obtained in section 5.3, we arrive at the following attractive
geometrical conclusion:  we can arrange that all extra $U(1)$ factors except hypercharge
acquire a St\"uckelberg mass,  provided we can find  compact CY manifolds with a del
Pezzo 8 singularity, such that only $\alpha_4$ represents a trivial homology class.
Requiring non-triviality of all other 2-cycles except $\alpha_4$ not only helps with
eliminating the extra $U(1)$'s, but also keeps a maximal number of gauge invariant couplings
in play as dynamically tunable moduli of the compact geometry.
In particular, to accommodate the construction of the SM quiver theory of fig 3,
the complex structure moduli of the compact CY threefold must allow for the formation of
an $A_2$ singularity within the del Pezzo 8 geometry.
%In the next section, we will present a general geometric  prescription for constructing

A general construction of a  compact CY embedding of the $dP_8$ singularity  with all the desired topological
properties was described in detail in \cite{Buican:2006sn}.

\begin{figure}[t]
\begin{center}
%\leavevmode\hbox{\epsfxsize=7.5cm \epsffile{East1.pdf}}\\[3mm]
\epsfig{figure=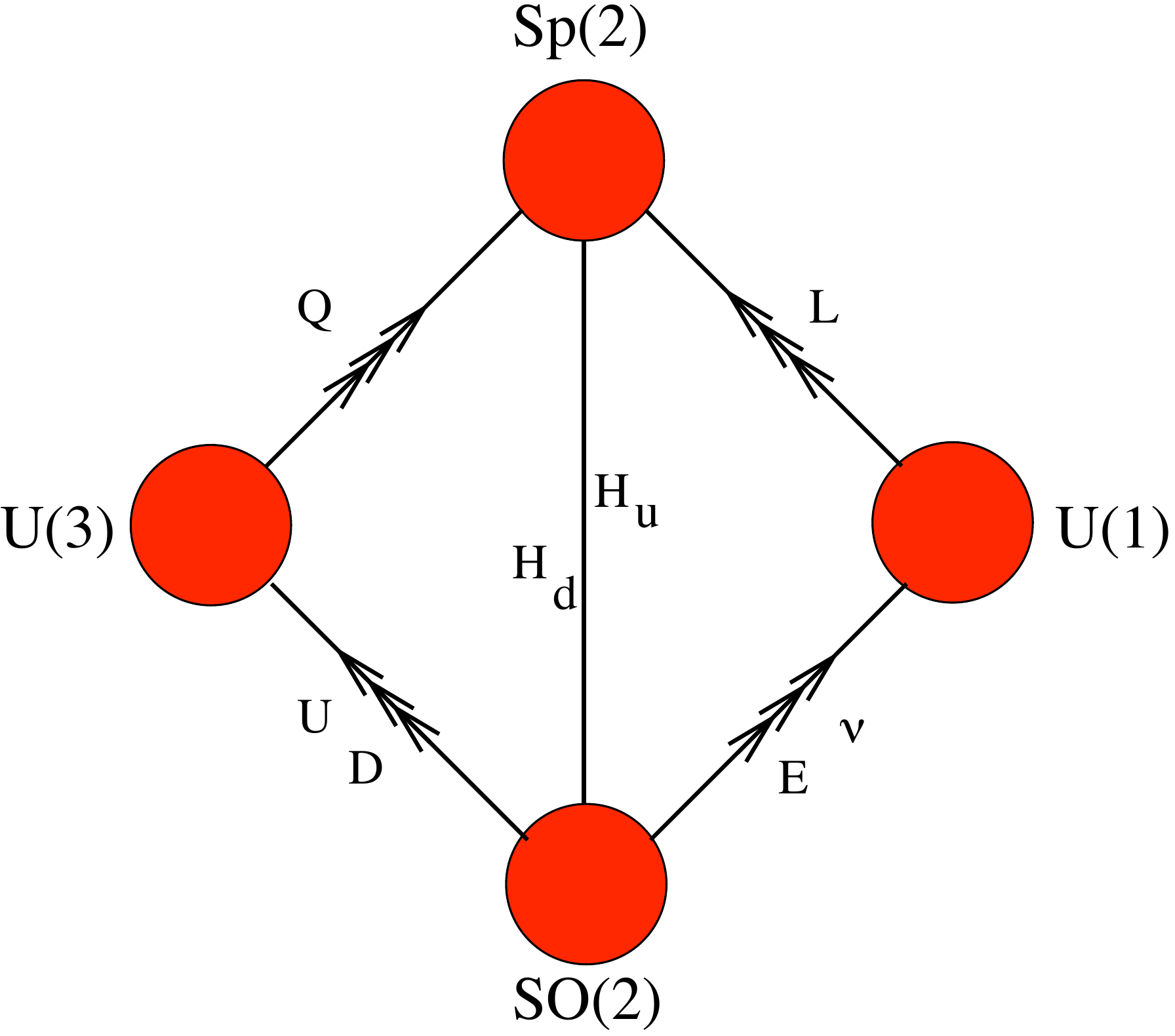,scale=0.36}
\vspace{-5mm}
\end{center}
\noindent
\caption{An MSSM-like quiver gauge theory, satisfying all rules for world-volume theories in an unoriented string models. A construction of this MSSM quiver theory via D-branes on a CY singularity
has recently been given in \cite{Wijnholt:2007vn}
}
%\label{default}
\vspace{-4mm}
\end{figure}

\bigskip

\section{Conclusions}

In these lecture notes we gave an introduction to some of
the concepts involved in constructing the SM-like gauge theories in
the systems of branes at singularities of CY manifolds.
After collecting some of the necessary technology, we then
outlined our general bottom-up
approach to string phenomenology. As an example of this approach,
we presented a concrete construction of an SM-like theory,  based on a single D3-brane near
a del Pezzo 8 singularity. We derived a simple topological condition
on the compact  embedding of the $dP_8$ singularity, such that only hypercharge survives
as the massless gauge symmetry.

The specific model based on the $dP_8$
singularity comes quite close to being realistic: it has the exact matter content and gauge
interactions of the SSM, except that it has a multitude of Higgs fields.
Furthermore,  supersymmetry is still unbroken.  We see no
a priori obstruction, however, to the existence of mechanisms that would lift all extra Higgses
from the low energy spectrum. SUSY
breaking terms may get generated via various  mechanisms: via fluxes, nearby anti-branes,
non-perturbative string physics, etc.  The structure of these terms is strongly
restricted by phenomenological constraints, such as the suppression of flavor changing
neutral currents.

The presence of the extra Higgs fields is dictated via the requirement (on all
D-brane constructions on orientable CY singularities)  that
each node should have an equal number of in- and out-going lines. To eliminate
this feature, it is natural to look for generalizations
among gauge theories on orientifolds of CY
singularities. Near orientifold planes, D-branes can support real gauge groups
like $SO(2N)$ or $Sp(N)$. With this generalization, one can draw a more minimal
quiver extension of the SM, with fewer Higgs fields. An example of such a
quiver is drawn in fig 5. It should be straightforward to find an orientifolded CY singularity
and fractional brane configuration that would reproduce this quiver.
The  extra $U(1)$ factors in  fig 5 can then be dealt with in a similar way
as in our $dP_8$  example.

\bigskip

\bigskip

\noindent
{\large \bf Acknowledgments}

These lecture notes are based on joint work with Matthew Buican, David Morrison, and
Martijn Wijnholt.  We thank the organizers of the 2006 Cargese School for their characteristic
hospitality. This work was supported by the National Science Foundation under
grants PHY-0243680 and DMS-0606578, by grant RFBR 06-02-17383 of the Russian Foundation of
Basic Research (D.M.).
% and by an NSF Graduate Research Fellowship (M.B.). D.M. would like to
%thank the IHES for hospitality and support when part of this work was done.
Any opinions,
findings, and conclusions or recommendations expressed in  this material are
those of the authors and do not necessarily reflect the views of the National Science
Foundation.

\end{document}